\documentstyle{article}
\def\Y{|}
\def\be{ \begin{equation}}
\def\ee{ \end{equation}}
\def\bea{ \begin{eqnarray}}
\def\eea{ \end{eqnarray}}
\newtheorem{theo}{Theorem}
\newtheorem{defn}[theo]{Definition}

\newtheorem{prop}[theo]{Proposition}
\newtheorem{coro}[theo]{Corollary}
\title{Construction of Field Algebras with Quantum Symmetry
       from Local Observables}
\author{Volker Schomerus \\
    Harvard University, Cambridge, MA 02138, U.S.A. }
\def\g{\phantom{W}}
\def\br{\overline}

\def\ba{\bea}
\def\ea{\eea}
\def\1{\iota }

\def\a{\alpha }
\def\b{\beta }
\def\c{\gamma }
\def\d{\delta }

\def\k{\kappa }
\newcommand{\vvert}[3]{(_{#1\ #2}^{\ \ #3})}
\def\Rep{{\cal R}{\it ep}}
\def\<>{\stackrel{\scriptscriptstyle < }{ \scriptscriptstyle >}}
\def\1{\iota }
\def\O{{\cal O}}
\def\T{{\cal T}}
\def\D{\Delta }
\def\H{{\cal H}}

\def\N{{\cal N}}
\def\A{{\cal A}}
\def\B{{\cal B}}
\def\G{{\cal G}}
\def\F{{\cal F}}
\def\U{{\cal U}}
\def\R{{\cal R}}

\def\C{{\cal C}}

\def\I{{\cal I}}
\def\K{{\cal K}}
\def\frc{ {\scriptstyle \frac12}}
\def\ti{\times }
\def\t{\tau}
\def\bt{\bar\tau}
\def\vp{\varphi }
\def\s{\sigma }
\def\tt{\tilde \tau}

\def\vac{ \Y 0 \rangle}
\def\inv{ \Y e_0 \rangle}
\def\vni{\langle e_0  \Y }
\def\nn{\nonumber}

\renewcommand{\!}{\hspace*{-2.5mm}}
\newcommand{\ew}{\hspace*{-2mm}}
\newcommand{\Fus}[6]{F_{{\scriptstyle #1} {\scriptstyle #2}}
                     \hspace*{-.5mm} [ \ew \begin{array}{ll}
                     {\scriptstyle #3 } \! & {\scriptstyle #4 }\!
                     \\[-2mm] {\scriptstyle #5}\! &{\scriptstyle #6}
                     \ew \end{array} ]}
\newcommand{\Br}[6]{B_{{\scriptstyle #1} {\scriptstyle #2}}
                    \hspace*{-.5mm} [ \ew \begin{array}{ll}
                    {\scriptstyle #3 }\! & {\scriptstyle
                    #4 }\! \\[-2mm] {\scriptstyle #5}\!
                    &{\scriptstyle #6}\ew \end{array} ]}
\newcommand{\Oe}{\Omega \vvert}
\newcommand{\OS}{\omega \vvert}
\newcommand{\CG}[6]{ [ \ew \begin{array}{lll}  {\scriptstyle #1} \!
                    & {\scriptstyle #2} \! & {\scriptstyle #3} \ew
                    \\[-2mm] {\scriptstyle #4} \! & {\scriptstyle #5}
                    \! & {\scriptstyle #6} \ew  \end{array}  ]}
\newcommand{\FS}[6]{\Phi_{{\scriptstyle #1} {\scriptstyle #2}}
                    \hspace*{-.5mm}  [ \ew \begin{array}{ll}
                    {\scriptstyle #3 }\! & {\scriptstyle #4 }\!
                    \\[-2mm] {\scriptstyle #5}\! &{\scriptstyle #6}
                    \ew \end{array} ]}
\newcommand{\pr}[1]{\ _{#1}}
\def\e{\epsilon}
\def\S{{\cal S}}
\def\o{\otimes }
\def\bo{\mbox{\,\raisebox{-0.65mm}{$\Box$} \hspace{-4.7mm}
${\scriptstyle\times}$ \/}}
\begin{document}
\begin{titlepage}
\maketitle \thispagestyle{empty}
\begin{abstract}
It has been discussed earlier that ( weak quasi-) quantum groups
allow for conventional interpretation as internal symmetries in
local quantum theory. From general arguments and explicit examples
their consistency with (braid-) statistics
and locality was established. This work addresses to  the
reconstruction of quantum symmetries and algebras of
field operators.
For every algebra $\A$ of observables satisfying certain standard
assumptions, an appropriate quantum symmetry is found.
Field operators are obtained which act on a positive definite Hilbert
space of states and transform covariantly under the quantum symmetry.
As a substitute for Bose/Fermi (anti-) commutation relations,
these fields are demonstrated to obey local braid relation.
\end{abstract}
\end{titlepage}

\tableofcontents

\section{Introduction}
\setcounter{equation}{0}

In classical mechanics, symmetries
are described by {\em groups} of  transformations
acting on a phase space. In view of the predominant role group
symmetries played in classical mechanics, it was natural to introduce
them also into quantum theory.
In fact, this was done soon after the discovery of quantum
mechanics and group symmetries turned out to be an important
tool to obtain  predictions of quantum theory \cite{Wig}.

Since the late seventies much work was done
to investigate quantum mechanical models on a two ore three
dimensional space-time.  Solving models of 1+1 dimensional
quantum field theory, Sklyanin, Takhtadzhyan and Faddeev
revealed a new algebraic structure that was called quantum
algebra \cite{FST,Fad}. In the more axiomatic treatment
of Drinfel`d and Jimbo \cite{Dri1,Jim} it appeared as a special class
of Hopf algebras \cite{Abe} and consequently as a generalization
of groups. The new name {\em quantum group} was used henceforth.
To cut a long story short we just mention that signs of
quantum group symmetries have been found in many quantum
mechanical models such as integrable spin chains
(e.g. \cite{PaSa,SaZu}),  rational conformal
quantum field theories (e.g. \cite{AGS,MoRe,MSI,HPT}),
and massive integrable models (e.g. \cite{ReSm}).
The process of generalization
continued. In 1989 Drinfel`d introduced quasi quantum groups
to be able to ``twist" quantum groups. Non-trivial examples
of quasi quantum groups were constructed  and
shown to be related to orbifold models \cite{DPR}.
For reasons that remain to be discussed later, all models
which exhibit generalized  {\em quantum symmetries}
are defined on a low dimensional space time.

To begin with, let us list the most important
algebraic structures which are common to all known examples
of quantum symmetries. The basic structure is an associative
*-algebra $\G^*$ (generalization of the group algebra).
Representations of $\G^*$ give rise to an action of the
quantum symmetry on vector spaces. Unitarity of the representation
$\t$ on a Hilbert space $V$ means that
$\t(\xi)^* = \t (\xi^*)$ for all $\xi \in \G^*$.
Among the representations of a quantum symmetry,
one can always find a ``trivial''
one--dimensional representation $\e$. Moreover, tensor
products of representations can be formed.
The tensor product of two representations
$\t,\t'$ on $V,V'$ is a representation
$\t \bo \t'$ on $V\o V'$. It is furnished by
a homomorphism  $\D: \G^* \to \G^* \o \G^*$
(a ``co-product'' of $\G^*$) according to the
formula
$$ (\t \bo \t' )(\xi ) = (\t \o \t')\D (\xi ) \ \ . $$
This tensor product need not be associative
and commutative and it may involve truncation,
i.e. the representation $\t \bo \t'$ possibly vanishes
on a non-trivial subspace of $V\o V'$. Finally, a notion of
``contragredient'' representations is furnished
by a suitable anti-automorphism $\S : \G^* \mapsto \G^*$ (an
``antipode'' of $\G^*$). We use the name ``bi-*-algebra
with antipode'' for such an algebraic structure. The precise
definition is given below in definition 3. The special case
where $\D(e) = e \o e$ and the co-product is co-associative
\footnote{co-associativity
is the property $(\D \o id)\D (\xi) = (id \o \D)\D(\xi)$ of the
co-product.} is called a ``Hopf-*-algebra''.

We adopt the framework of second quantized quantum
mechanics, so that there is a Hilbert space
$\H$ of physical states which is generated from
a unique ground state $\vac$ by application of
a set of field operators $\Psi^I_i (x,t)$. Superscripts
$I,J,K,\dots $ distinguish between field multiplets
while subscripts label members of the multiplets. A
quantum mechanical system is said to possess
a quantum symmetry $\G^*$, if
the Hilbert space $\H$ carries an unitary
representation $\U$ of $\G^*$,
such that the
ground state $\vac $ is invariant and field
operators $\Psi^I_i (x,t)$ transform covariantly.
Invariance of the ground state means
that $\vac$ transforms according to the
trivial one--dimensional representation $\e$
of $\G^*$. The formulation of the transformation
law of field operators involves the tensor product
of representations of $\G^*$. More precisely, a field
multiplet $\Psi^I_i(x,t)$ is said to transform covariantly
according to the finite dimensional representation
$\t^I$ of $\G^*$, if
$$ \U(\xi) \Psi^I_i(x,t) = \Psi^I_j (x,t) (\t^I \bo \U)_{ji} (\xi) $$
holds for all $\xi \in \G^*$ \cite{BMT2}.
The symmetry transformations
considered here, do not act on the space time argument of
the field operators, in other words: they are internal
symmetries. We give a more comprehensive explanation
of the notion of quantum symmetry in the next
section.

In second quantized quantum theory, Bose and Fermi statistics are
implemented through local commutation or anticommutation relations of
field operators which create particles,
\be
  \Psi^I_i(x, t)\Psi^J_j(y, t)
  =  \pm  \Psi^J_j(y, t)
  \Psi^I_i(x, t)  \ \  \mbox{ for } x \not= y \ \ .
\ee
Consistency of a symmetry with Bose/Fermi statistics requires that
this relation should be preserved by a symmetry transformation.
This is indeed true for internal symmetry groups.

In two and less space dimensions, Bose/Fermi statistics is not the
most general possibility, but braid group statistics can also occur.
Fr\"ohlich  proposed that local (anti-) commutation relations of
fields should be be replaced by {\em local braid relations}.
\be
 \Psi^I_i(x,t)\Psi^J_j(y,t)
  = \omega^{IJ}  \Psi^J_l(y,t)
  \Psi^I_k(x,t) \hat\R^{IJ>}_{kl,ij}   \label{braidrel}
\ee
if $x>y$ for some ordering of space coordinates
and $\omega^{IJ}$ are phase factors. Originally, $R^{IJ>}_{kl,ij}$
were proposed to be complex numbers.

The main question is, whether such local braid relations
can be consistent with the transformation law under some
non-trivial quantum symmetry. The answer is affirmative.
It will be seen, however, that braid matrices $\hat \R $
with entries in
$\G^*$ should be admitted.

In general, local braid relations of this more general type
can be consistent with non-trivial quantum symmetries, provided
the tensor product of representations is associative
and commutative at least up to equivalence.
\bea
(\t^I \bo \t^J) &\cong& (\t^J \bo \t^I) \label{repcom} \\[1mm]
(\t^I \bo \t^J) \bo \t^K & \cong & \t^I \bo
                          (\t^J \bo \t^K) \ \ .\label{repass}
\eea
The precise formulation of this requirement will involve
an element $ R \in \G^*\o \G^*$
and a ''re-associator'' $\vp \in \G^* \o \G^*\o \G^*$
with certain properties. They will furnish invertible
intertwiners $R^{IJ} = (\t^I \o \t^J)(R)$
and $\vp^{IJK}= (\t^I \o \t^J \o \t^K)( \vp)$ between the
representations on the right and left hand side of eq.
(\ref{repcom},\ref{repass}). When
$\vp = \sum_{\s } \vp^1_{\s }\o \vp^2_{\s }\o \vp^3_{\s}$
one introduces
$ \vp_{213} = \sum_{\s}\vp^2_{\s}\o \vp^1_{\s}\o \vp^3_{\s}$.
In this notation, consistent braid matrices are given by
\be       \label{consR}
\hat{\R}^{IJ>}_{k l ,i j}= (\t^{I}_{ki}\o \t^{J}_{lj}\o \U)
  (\vp_{213}(R\o e)\vp^{-1})\in \U(\G^*)\ \ .
\ee
Bi-*-algebras with antipode $\S$,
re-associator $\vp$ and $R$-element $R$
are called {\em weak quasi quantum groups}. They were introduced
in \cite{MSIII} as a generalization of Drinfeld's quasi quantum
groups \cite{Dri2}. A precise definition will be given in section 4.
Consistency of weak quasi quantum group symmetries with
local braid relations is discussed in
more detail in section 5.1.

It is shown in \cite{MSIV} that the chiral Ising model -- i.e.
the conformal quantum field theory in which observables are
generated by a Virasoro algebra with central charge
$c=\frac12 $ -- provides an example, with the truncated
quantum group algebra $U^T_q(sl_2),q=\pm i, $ as a symmetry
\footnote{$U_q^T(sl_2)$ is a family of weak quasi quantum groups
which is obtained from $U_q(sl_2)$ by a process of truncation
\cite{MSIII}.}. To the best of our knowledge,
this was the first time that the consistency
of non-abelian local braid relations (\ref{braidrel}) has been
demonstrated through the construction of a model. Originally
it had been proposed that minimal conformal models have
quantum groups as symmetries \cite{AGS,PaSa},
but this identification is not quite satisfactory, because the local
braid relations, which should come with the symmetry, are not
satisfied \cite{MSI}.

The main result of this work states that the picture we obtained
from studying this example is generic.
The notion of fusion rules of superselection sectors will
be explained in detail in section 3.1. Basically, superselection
sectors are inequivalent *-representations of the algebra of observables
in quantum mechaniocs. Under standard assumptions, a tensor products of
such representations can be defined and decomposed into irreducibles,
with multiplicities $N^{IJ}_K$ ("fusion rules").
The algebra of observables $\A $ has subalgebras $\A(\O )$
which correspond to measurements in the bounded space time domain $\O $.
It suffices to consider double cones $\O $, i.e. nonvanishing
                                                 intersections
of the interior of forward and backward light cones. These algebras
form a net, i.e. there is
an inclusion preserving map $\O \mapsto
\A(\O)$, where $\O$ is an element in the set $\K$ of open double
cones in spacetime.

\begin{theo} {\em (QFT Reconstruction theorem)}
\label{QFTreconst}
Let $\A$ be a net of local observables which satisfies the
standard assumptions (Haag-Kastler axioms, Haag-duality,
locally generated sectors, finite statistics). Suppose that
there is a bi-*-algebra with antipode
such that the multiplicities
in the Clebsch-Gordon decomposition agree with the fusion
rules $N^{IJ}_K$ defined by the superselection structure of $\A$.
Then one can construct a {\em complete local field system
$(\F,\G^*,\H,\pi)$ with quantum symmetry}.
In detail this means the following.
\begin{enumerate}
\item There exists a net $\F$ of local quantum fields. The
      algebras $\F(\O)$ which $\F$
      assigns to every open double cone $\O$ in spacetime come
      equipped with a conjugation  $\overline{\g}$. The product in
      $\F(\O)$ is denoted by $\cdot$.
      The product is not necessarily associative, but it enjoys
      quasiassociativity properties as stated in theorem \ref{QASSF}
      below.
\item $\G^*$ is a weak quasi quantum group with re-associator
      $\vp$ and $R$-matrix $R$. Elements $\xi \in \G^*$ act on
      $\F(\O)$ as generalized derivations. The algebra $\A(\O)$
      of observables  in $\O$ is the fixed-point algebra
      with respect to this action.
\item Quantum fields localized in relatively spacelike regions of
      spacetime satisfy local braid relations. If
      $\psi_\a \in \F(\O_1)$, $\psi'_\b \in \F(\O_2)$ transform
      according to irreducible representations $\t \cong \t^J$
      and  $\t' \cong \t^K$ of $\G^*$, they read
    \be
      \psi_\a \cdot \psi'_\b = \omega^{JK} \psi'_\c \cdot \psi_\d
      (\tau_{\d\a} \o \tau'_{\c\b})(R)
    \ee
      whenever $\O_1 > \O_2$. Here $\omega^{IJ}$ are certain phase
      factors depending on the equivalence classes of $\t,\t'$.
      Fields are local relative to the observables, i.e. $\F(\O_1)$
      and $\A(\O_2)$ commute elementwise whenever $\O_1$ and $\O_2$
      are spacelike separated.
\item There is a Hilbert space $\H$ which carries a unitary
      representation $\U$ of $\G^*$ and a faithful representation
      $\pi$ of $\F$. When $\pi$ is restricted to $\A$, it determines
      a reducible representation of $\A$ on $\H$. $\H$ decomposes
      into a direct sum of ``superselection sectors'' $\H^J$
      (irreducible representations of $\A$) each with some finite
      nonzero multiplicity $\delta_J$.
$$ \H = \sum_J \H^J\otimes V^J \ , \  \ \ dim V^J = \delta^J \ . $$
      $V^J$ are representation spaces for $\G^*$.
      States within the ``vacuum
      sector'' $\H^0$ are $\G^*$-invariant. $\H^0$ appears with
      a multiplicity $\delta_0= 1$ and is cyclic for $\pi(\F(\O))$.
\item Suppose that $\psi_\a, \psi_\b$  transform according to
      representations $\t,\t'$ of $\G^*$ and denote their
      ``representation operators'' $\pi(\psi_\a),\pi(\psi'_\b)$
      (``field operators'') as
      $\Psi_\a, \Psi'_\b$. Then the product $\psi_\a \cdot \psi'_\b$
      is represented by
    \be \pi(\psi_\a\cdot \psi'_\b) = \sum \Psi_\c \Psi'_\d
      (\t_{\c\a} \o \t'_{\d\b} \o \U)(\vp) \ \ .
    \ee
      It follows in particular that field operators obey
      local braid relations of the form (\ref{braidrel})
      with a (not necessarily numerical) $\R$-matrix.
\end{enumerate}
\end{theo}

The braid relations hold for all field operators, including composites,
i.e. the braid relations pass to products and conjugates. The
field operators satisfy quantum symmetric operator product expansions,
see theorem \ref{OPE}.

\begin{theo} {\em (existence of the quantum symmetry)} An
appropriate bi-*-algebra $\G^*$ with antipode as required
by the hypothesis of the preceding theorem
exists, if the theory defined by $\A$
is rational, i.e. has only a finite number of superselection
sectors.
\end{theo}

The above theorems and their proofs are expressed in the language
of algebraic quantum field theory \cite{Haa}.
Some basic ideas from the algebraic theory of superselection
sectors are reviewed in section 3. Covariant field operators
will be constructed in section 3.3.. Local braid relations
are established in section 5.2.
To discuss the locality properties of field operators
some mathematical results on symmetry
algebras with re-associator $\vp$ and $R$-element $R$ are needed.
In order not to clutter the presentation in section 5,
this mathematical background will be anticipated in section
4. Readers who do not want to enter the discussion of algebraic
quantum field theory but have some basic knowledge in
two-dimensional conformal  quantum field
theory may skip section 3.1,3.2.. Besides of providing
a more rigorous mathematical basis, arguments from algebraic
theory of superselection sectors are
used to demonstrate the universality
of ``vertex operators'' and their properties.
Indeed all related features familiar from two-dimensional
conformal theories extend to a wide class of two- and
three-dimensional models (without conformal symmetry).

Internal symmetries recently
led to a classification of quantum field theories with
permutation group statistics \cite{DoRo}.
Doplicher and Roberts proved that a unique compact
symmetry group $G$ can be associated with every
higher dimensional quantum field theory.
Moreover, the quantum field on which the symmetry
transformations act, commute or anticommute for spacelike
separations. Theorem \ref{QFTreconst} generalizes the
construction of Doplicher and Roberts
to quantum field theories with
braid group statistics. Some remarks on uniqueness, which
fails to hold in the setting of theorem \ref{QFTreconst}, are
included in the last section. For related work see
Fr\"ohlich and Kerler \cite{Ker1,KerB},
Majid \cite{Maj2}, Rehren \cite{Reh1,Reh2} and Todorov et. al.
\cite{HST,HPT}.
\section{The Notion of Quantum Symmetry}\setcounter{equation}{0}
Let us begin with a short review on group symmetries in quantum
mechanics.
Consider some quantum mechanical system $(\H,\{ \Psi\} ,\vac,H)$
                     within a second
quantized formalism. The Hilbert space $\H$ of physical states
should contain a unique ground state $\vac$ with respect to the
Hamiltonian $H$. We assume that $\H$ is generated from $\vac$ by
multiplets
of  field
operators $\Psi^I_i (x,t)$ (I labels multiplets, i labels fields in
the multiplet I) which
create particles or excitations.

A compact group $G$ is called (internal) symmetry of this system
if there is a unitary representation $\U: G \to \B(\H)$
such that the ground state $\vac$ and the Hamiltonian $H$ are
invariant and field operators $\Psi^I_i$
 transform covariantly according
to the representation $\tau^I$ of $G$. To state these requirements
in mathematical terms,
let us recall two notions from the representation theory
of groups. Every group has a trivial representation $\e_G: G \to
{\bf C}$ defined by $\e_G (\xi) = 1$ for all $\xi \in G$. The
tensor product $\bo$ of representations $\tau, \tau'$ is given by
\be (\tau \bo \tau')_{kl,ij} (\xi) = \tau_{ki}(\xi) \tau'_{lj}(\xi)
\ \ \ \mbox{ for all } \ \ \ \xi \in G \ \ . \label{gtensprod}\ee
If we set $\xi^* = \xi^{-1}$, unitarity of $\U$ asserts  $\U(\xi)^*=
\U(\xi)^{-1} = \U(\xi^{-1}) = \U(\xi^*)$. Invariance
of the ground state $\vac$ can be expressed as
\be   \U(\xi)\vac = \vac  = \vac \e_G(\xi) \ \
        \mbox{ for all } \xi  \in G\ \ \ . \ee
We say that $\Psi^I_i(x,t )$ transforms covariantly according to
the representation $\tau^I$ of $G$, if for all $\xi \in G$
\be
\U(\xi ) \Psi^I_i(x,t )  =   \Psi^I_j(x,t ) \tau^I_{ji} (\xi )
\U(\xi)  = \Psi^I_j (x,t) (\tau^I \bo \U)_{ji}(\xi)
 \ .
\ee
Since we concentrate on internal symmetries, there
is no action on the space-time variable of the field.
For this reason we will often neglect
to write arguments  $(x,t)$ explicitly.
Adjoint field operators $\Psi^{I*}_i$ transform covariantly
according to the ``conjugate'' representation $\bt^I(\xi) \equiv
(\ ^t\t^I(\xi))^*$, for all $\xi \in G$ (here
and in the following $\ ^t$ denotes the
transpose of a matrix and $*$ refers to a scalar product on
the representation space of $\t^I$).
\be
\U(\xi ) \Psi^{I*}_i
  = \Psi^{I*}_j  (\bt^I \bo \U)_{ji}(\xi)
\ .
\ee

In conclusion, the formulation of symmetry in quantum theory
involves a conjugation $*$ to express unitarity, a trivial
representation $\e$ to state invariance and a tensor product $\bo$
of representations to write down the covariance law.
The mathematical structure  behind these notions
is known as bi-*-algebra with antipode.

\begin{defn}{\em (bi-*-algebra with antipode)}\label{BIAA}
A (not necessarily co-associative) bi-*-algebra
($\G^*,\D,\e,*$) is furnished by
an associative  *-algebra $\G^* $,
with unit $e$ and $\D: \G^* \to \G^* \o \G^*$ ({\em co-product}),
$\e: \G^* \to {\bf C}$ ({\em co-unit}) which are *-homomorphisms.
$\D $ is not required to be unit-preserving, i.e. $\D (e)\not= e\o e$
is permitted.
The co-product $\D$ and the co-unit $\e$ must satisfy
\be
 (\e \o id) \D = id = (id \o \e) \D  \ \ \ .\label{edrel}  \ee
The bi-*-algebra has an antipode if there exists a {\bf C}-linear
antiautomorphism
          $\S: \G^* \mapsto \G^*$ ({\em antipode})
such that
\be \sum_\s \S(\xi^1_\s) \a \xi^2_\s = \a \e(\xi) \ \ \ \ , \ \ \ \
\sum_\s \xi^1_\s \b \S(\xi^2_\s)= \b \e(\xi) \ \
\label{antip} \ee
hold for two fixed (independent of $\xi$) elements $\a,\b \in \G^*$
and
\be \S(\xi)^*  = \S^{-1}(\xi^*) \ \ \ \ , \ \ \ \
\b = \S(\a^*)  \label{staranti}\ee
If $\D (e) = e\otimes e $ and if the co-product is co-associative,
viz. $(\D \o id )\D = (id \o \D )\D $ then
$\G^*$ is called a `` Hopf-*-algebra''.
The bi-*-algebra is called semisimple if $\G^*$
is a sum of full matrix algebras.
\end{defn}

Here $\xi^i_\s$ are defined by the expansion $\D(\xi) = \sum_\s
\xi^1_\s \o \xi^2_\s$ of the co-product $\D$. Consistency of the
antipode with the $*$-operation is achieved by eqs.(\ref{staranti}).

For $\D$ the notion of a *-homomorphism involves a definition of
$*$ on $\G^* \o \G^*$, which is not unique (cf. \cite{MSIII}) since
there are two possibilities to define a *-operation on
$\G^{*\o_{N+1}}$
\bea \mbox{ (I) } \ \ (\xi \o \eta)^* &=& \xi^* \o \eta^*
                                     \label{conj1}         \\
\mbox{ (II) } \ \ (\xi \o \eta)^* &=& \eta^* \o \xi^*
                                     \label{conj2}    \eea
for all $\xi \in \G^*, \eta \in \G^{*\o_N}$.
Throughout this paper we will only consider conjugations of
type (I). Everything below can be worked out for conjugations
of type (II) ( cp. \cite{MSadd} for formulae).
Note that $\D (e) = e \o e$ is not assumed. For us, the
the homomorphism property of $\D$ means $\D(\eta \xi) =
\D (\eta ) \D (\xi)$ and implies only that
$\D( e)$ is a projector, i.e. $\D(e) \D(e) = \D(e)$.
When we want to stress that $\D$ is not unit preserving,
$(\G^*,\D,\e,*)$ will be called ``weak'' bi--*--algebra.
One should also note  the freedom in the choice of $\a,\b$.
in the definition of the antipode.
For a given antipode $\S$, relations (\ref{antip},\ref{staranti})
are invariant under the transformation $\a \mapsto \zeta \a$
with an arbitrary element $\zeta$ in the center of $\G^*$.

The co-product $\D$ determines a tensor product $\tau \bo \tau'$
for representations $\tau,\tau'$ of $\G^*$
\be  \label{tensprod} (\tau \bo \tau')(\xi) = (\tau \o \tau') (\D(\xi))
\ \ \mbox{for all } \ \xi \in \G^* \ \ . \ee
With respect to this tensor product of representations, the co-unit
$\e$ furnishes a trivial one--dimensional representation. Triviality
refers to the property
$\e \bo \tau = \tau = \tau \bo \e $ for all
representations $\tau$ of $\G^*$, which follows from (\ref{edrel}).
If $\D (e) \neq e \o e$ the tensor product of representations
is truncated. This means that it is zero on a non-trivial subspace
of the tensor product of representation spaces.

A ``contragredient'' representation $\tt$ of the representation $\t$
of $\G^*$ can be defined with the help of the antipode $\S$.
\be   \tt (\xi) \equiv \ ^t\t(\S^{-1}(\xi)) \ \ \mbox{ for all }
   \ \ \xi \in \G^* \ \ .
\ee
Relations (\ref{antip}) assert that the tensor products
$\tt \bo \t$ and $\t \bo \tt$ contain the trivial representation
$\e$ as a subrepresentation. It is this feature which motivates the
name ``contragredient'' representation. We introduce
the symbol $\bt$ to denote the representation of $\G^*$
obtained as
\be  \label{conjrep}
\bt (\xi) \equiv  (\ ^t\t (\S^{-1}(\xi^*)))^*   \ \ \mbox{ for all }
\ \ \xi \in \G^*     \ \ .
\ee
Since $\bt$ will turn out to  determine the transformation
law of adjoints, we refer to $\bt$ as ``conjugate'' representation.
If the representation $\t$ is unitary, i.e.
 $\t(\xi^*) = (\t(\xi))^*$ for all $\xi \in \G^*$, the
definition (\ref{conjrep}) assumes the simpler form $\bt(\xi) =
\ ^t\t(\S(\xi))$.

Let us explain how to abstract a bi-*-algebrawith antipode
from the representation theory of the compact group $G$.
In this case,
$\G^*$ should denote the group algebra of the compact gauge group $G$,
i.e. a space  of  ``linear combinations''
of elements in $G$. All (anti-) homomorphisms of
the group $G$ can be uniquely extended to algebra-homomorphisms of
the group algebra $\G^*$. Consequently it suffices
to fix $\D_G, \e_G,*,\S_G$ on elements $\xi$ in the group $G$.
$\e_G, *$ have been defined above and comparison of (\ref{tensprod})
with (\ref{gtensprod}) yields
\be  \Delta_G (\xi) = \xi \o \xi \ \ \ \mbox{ for all } \xi \in G
 \ \ \ .  \label{DeltaG} \ee
Similarly we obtain the expression $\S_G(\xi) = \xi^{-1}$ for the
antipode.
Since $\D_G(e) =e \o e$,  the co-product $\D_G$
is unit preserving. Assuming that the
action of * on $\G^* \o \G^*$ is specified by $(\xi \o \eta)^*
= \xi^* \o \eta^* $, ($\G^*,\D_G,\e_G,*,\S_G$)
is easily shown to satisfy
all assumptions listed above. In this sense, group algebras
are only special
examples of bi-*-algebras with antipode.
\begin{defn} {\em (quantum symmetry) \cite{MSIII}}
A *-algebra $\G^*$ with co-product $\D$ and co-unit $\e$
is called {\em quantum symmetry} of
the system $(\H,\{ \Psi \} ,\vac,H)$,
if there exits a representation
$$\U : \G^* \mapsto \B (\H) \ \ \ \ \ \ \ \mbox{ such that }$$
i) $\U$ is unitary in the sense that $\U (\xi ^*) = \U(\xi)^*$ for
all $\xi \in \G^*$. \\
ii) the Hamiltonian H and the ground state $\vac$ are invariant, i.e.
\be   [H,\U (\xi)] = 0  \ \ \ \ , \ \ \ \U (\xi)\vac = \vac \e(\xi)
\ \ \mbox{ for all } \ \ \xi \in \G^* \ \ .\ee
iii) the field operators $\Psi^I_i (x,t)$ transform covariantly with
respect to the representation $\tau^I$ of $\G^*$, i.e.
\bea
\U(\xi ) \Psi^I_i(x,t)  & = & \sum_p \Psi^I_j(x,t) (\tau^I\bo
\U)_{ji} (\xi) \label{covtrans} \\
            & = & \sum_p \Psi^I_j(x,t) \tau^I_{ji} (\xi^1_p )
\U(\xi^2_p ) \  ,                 \nn
\mbox{ if } \Delta(\xi )  =  \sum_p \xi^1_p \o \xi^2_p \ .
\eea \end{defn}

The covariant transformation law (\ref{covtrans})
tells us how to shift representation
operators $\U(\xi )$ through fields from left to right. Together
with the invariance of the ground state $\vac$ it determines
the transformation properties of states. We demonstrate
this for  the 1-excitation states.
$$
\U(\xi ) \Psi^I_i \vac   =  \sum \Psi^I_j \tau^I_{ji}
(\xi^1_p) \U(\xi^2_p) \vac   =
\sum \Psi^I_j \tau^I_{ji}(\xi^1_p) \epsilon(\xi^2_p) \vac
 =  \Psi^I_j \vac \tau^I_{ji} (\xi )\ \ .
$$
The transformation law of
higher excitations can be found along the same lines. As a result
one finds that
they transform according to some tensor product of representations
$\tau^I$.
$$  \U (\xi) \Psi^{I_1}_{i_1} \dots \Psi^{I_n}_{i_n} \vac =
\Psi^{I_1}_{j_1} \dots \Psi^{I_n}_{j_n} \vac (\tau^{I_1} \bo
(\tau^{I_2}\bo (\dots  \bo \tau^{I_n})))_{j_1 \dots j_n,
i_1 \dots i_n}(\xi)\ \ . $$
The brackets  in the tensor product of representations on the
right hand side are necessary since the tensor product
(\ref{tensprod}) need not be associative.
We will come back to this
point later.

We would like to derive a transformation law for adjoint fields
$\Psi^{I*}_i$ as it was done for group symmetries. However, the
existence of an antipode $\S$ does not suffice for this purpose, even
though it furnished an appropriate notion
of contragredient and conjugate
representations.
Let us assume in addition that there exists an
element $\vp \in \G^* \o \G^* \o \G^*$ such that
\be
\vp (\D \o id) \D(\xi) = (id \o \D) \D(\xi) \vp \ \ \mbox{ for all }
\ \ \xi \in \G^*\ \ .
\label{vpiint} \ee
We will give a detailed discussion of such elements $\vp$ below.
They are called re-associators. Here we use
the existence of $\vp = \sum \vp^1_\s \o \vp^2_\s
\o \vp^3_\s$ to infer
transformation properties of a covariant adjoint of field operators.
The covariant adjoint can be regarded
as an analog of the Lorentz covariant adjoint $\br\Psi = \Psi^*\gamma_0$
of a Dirac spinor field.
\begin{defn}{\em (covariant adjoint)}\label{COVADJ}
\bea   \label{covad}
\br\Psi^I_i &\equiv& (\Psi^I_j (\t^I_{ji} \o \U)(w))^* =
\t^I_{ij}(w^{1*}_\s) \U(w^{2*}_\s) \Psi^{I*}_j    \\[1mm]
\mbox{ with } \ \ w & = & \sum w^1_\s \o w^2_\s \equiv
\sum \vp^2_\s \S^{-1}(\vp^1_\s \b ) \o \vp^3_\s \nn
\eea
\end{defn}
The second equality in the definition (\ref{covad}) of the
covariant adjoint holds only if $\t^I$ is unitary.
\begin{prop} The covariant adjoint
transforms covariantly according to the representation $\bt^I$
of $\G^*$.
\be
\U(\xi) \br\Psi^I_i = \br\Psi^I_j (\bt^I \bo \U )_{ji} (\xi) \ \ .
\label{adtrans} \ee
\end{prop}
{\sc Proof:} From relations (\ref{antip}) and (\ref{vpiint})
one obtains
$$ w (e \o \xi) = \sum \D(\xi^2_\s) w (\xi^1_\s \o e) $$
for all $\xi \in \G^*$. It follows that
$$ \Psi^I_j (\t^I_{ji} \o \U)(w) \U(\xi) =
(\tt^I_{ik}\o \U)(\D(\xi)) \Psi^I_k (\t^I_{kj} \o \U)(w)\ \ . $$
Taking the adjoint of this equation one derives the
formula (\ref{adtrans}).

One should notice that the properties of the antipode $\S$ (\ref{antip})
and the discussion of covariance for adjoint fields
differ from the earlier work (\cite{MSIII} f.).
The original treatment
turned out to provide an unnecessary restriction within
the class of interesting quantum symmetries. In addition many
crucial statements about covariant adjoints given below do
not hold within the setting of \cite{MSIII}.It was suggested
by Vecsernyes \cite{Vec} to use Drinfeld's definition
(\ref{antip}), and this is indeed appropriate as the
results reported here confirm.

\section{Algebraic Methods for Field Construction}
                                   \setcounter{equation}{0}
According to the laws of local relativistic quantum mechanics,
observables are selfadjoint operators acting in a Hilbert  space
${\cal H}$ of physical states.
The Hilbert space of physical states ${\cal H}$ may
decompose into orthogonal subspaces ${\cal H}^J$,
called {\it superselection sectors}, such that
observables $A$ do not make transitions between different
subspaces ${\cal H}^J$ \cite{WWW}.
Different sectors ${\cal H}^J$ carry inequivalent
irreducible positive energy representations
of the algebra ${\cal A}$ of observables $ A$,
possibly with some multiplicity \cite{HaKa}. Among the sectors
is the unique vacuum sector ${\cal H}^0$ which contains the
vacuum $\vac $ and appears with multiplicity 1.

When several superselection sectors exist, it is of interest
to construct {\em additional field operators} which make
transitions between superselection sectors so that the whole
Hilbert space of physical states is generated from the
vacuum by application of field operators. These fields
operators should commute with all observables when
their space--time arguments are spacelike localized.

By definition,
observables have to commute with the generators
of an internal gauge symmetry. A field operator which transforms
according to some non-trivial representation of an internal
gauge symmetry is necessarily non-observable,
i.e. it maps states in different superselection sectors into
each other. This implies that
states in different sectors possibly
transform according to inequivalent representations
of the internal symmetry.

According to the algebraic theory of
superselection sectors \cite{DHR1,DHR2},
non-observable fields are constructed
by adjoining localized endomorphisms $\rho$ to the algebra
$\A$ of observables. These ideas will be explained in some
detail below. For a comprehensive introduction we
recommend the recent book of Haag \cite{Haa} and
lectures by Roberts in \cite{KasB}.

\subsection{Observables and superselection sectors in
local quantum field theory}

In this section, $M_d$ is a d--dimensional space-time manifold with
a global causal structure. $\K$ will denote the set of
all double cones $\O$ (non void intersections of
forward and backward light cones) in $M_d$.
To be specific let us choose
$M_d$ to be the d--dimensional Minkowski space.
We will consider another
possibility at the end of this subsection.
Two subsets $\S_1,\S_2$ of $M_d$ are called spacelike separated, if
any two points $x_i \in \S_i$ are relatively spacelike.
The causal complement $\S'$ of a  $\S \subset M_d$ is
the set of all points spacelike separated from $\S$.
If d=2 and $\O$ a double cone, $\O'$ decomposes
into two disconnected components $\O_{\<>}$. We use the
notation $\O_1 \<> \O$ whenever $\O_1 \subset \O_{\<>}$.
In the following, $G$ is a group of global transformations
of $M_d$ which respect the causal structure. It is supposed
to contain space-time translations. Elements $g \in G$
transform regions $\O \in \K$ to  $g\O$.

Consider a set of observable local quantum fields $\phi_b(x)$
on the space-time $M_d$. One might for example think of
an energy--momentum tensor $T(x)$ or a family of real currents
$J_a(x)$. They are described by a commutator which
vanishes for relatively spacelike arguments.
According to the Wightman theory \cite{GaWi,StWi},
fields are operator valued distributions. So they should
be evaluated on real test functions $f: M_d \to ${\bf R}
to obtain (unbounded) operators on the Hilbert space of
physical states.
$$ \phi_b(f) = \int_{M_d} dx \phi_b(x) f(x)\ \ . $$
Regarded as operators on the vacuum sector $\H^0$, these
smeared out fields ''generate'' the algebra of observables.
We define an algebra $\A (\O)$ of observables localized in $\O$
to be the von Neumann (weakly closed *--) algebra
generated by all bounded functions of the operators
$\phi_a(f)$ with {\it supp}$(f) \in \O$. Properties of
the set of observable fields $\phi_b(x)$ can be translated
into properties of the family $\A(\O)$ they generate.
We would like to mention that in this step one has to be aware of
certain subtleties which might be overlooked at a first glance.
In general, domain problems of the unbounded operators
$\phi_a(f)$ spoil local commutativity $[\Phi(f_1),\Phi(f_2)] = 0$
of bounded functions $\Phi$  in $\phi(f_i)$ for spacelike separated
support of $f_i$.
For a review on the status of
these problems see \cite{Yng} and references therein.

At least in many applications to concrete models, we
are led (cf. \cite{Haa} for a thorough discussion)
to consider a family of von
Neumann algebras $\A (\O)_{\O \in \K}
\subset \B(\H^0)$ in a Hilbert space $\H^0$ which
satisfies the following basic (Haag-Kastler)
axioms \cite{HaKa}.
\begin{enumerate}
\item
It is isotonic, i.e. $\O_1 \subset \O_2 \Rightarrow
\A(\O_1) \subset \A(\O_2)$.
\item
Einstein causality (or locality) is satisfied, which means
that $\A(\O_1) \subset \A(\O_2)'$ if $\O_1 \subset \O_2'$. Here
$\A(\O_2)'$ is the set of all bounded operators in $\H^0$ which
commute with $\A(\O_2)$.
\item
There is a strongly continuous unitary representation $U_0$ of G
in $\H^0$ which implements automorphisms $\a_g: \A(\O) \to \A(g(\O))$
defined by
\be
\a_g(A) = U_0(g) A U_0(g^{-1})  \ \ \ \mbox{for every } g \in G\ \ .
\ee
The generators of the translation subgroup
should have their spectrum in the closed forward light cone.
We will refer to these properties as covariance and spectrum
condition respectively.
\item
There is an (up to a phase) unique vector $\vac  \in \H^0$
which is invariant under the action of G, i.e. $U_0(g) \vac= \vac$.
$\vac$ is cyclic for each $\A (\O)$.
\end{enumerate}
Algebraic quantum field theory starts from this algebraic
structure. Specific
properties of underlying Wightman fields (which often
exist in the applications but possibly not in general) are
not needed in the analysis. So we might take the family $\A$
of local observables instead of a set of Wightman fields
to define (the observable content of) the model.

All the properties of the family $\A$ stated above, reflect deep
physical principles (Einstein causality, covariance, etc.).
Later we will often need
another assumption on the structure
of ${\A (\O)}_{\O \in \K}$,
which cannot be justified on the same footing.
The family $\A$ is said to satisfy
{\em Haag duality} if
\be \A(\O) =  \A(\O')' \ \ \ \mbox{ for all }
\ \ \O \in \K \ \ .\label{Hdual}
\ee
Here $\A (\O')$ is defined to be the $C^*$-algebra
generated by the algebras $\A( \O_1)$ for $\O_1 \subset \O'$,
$\O_1 \in \K$. Haag duality can be regarded as a strong
version of Einstein causality, which asserts that
$\A(\O) \subset \A(\O')'$.
Generally speaking, breakdown of this
duality  indicates spontaneous breakdown of symmetry \cite{Rob}.
Bisognano and Wichmann have shown that for families $\A$ generated
by Wightman fields one can always pass to the bi-dual $\B (\O) :=
A(\O)')'$ which then satisfies duality \cite{BiWi1,BiWi2}.
We shall make some more specific remarks on
validity of (\ref{Hdual}) in conformal quantum
field theories below.

According to our introductory remarks, superselection sectors
carry irreducible
representations of the observables algebras
$\A(\O)_{\O \in \K}$.
A representation of $\A(\O)_{\O \in \K}$  is a family of representations
$\pi^{\O}, \O \in \K$ of $\A( \O )$ on some Hilbert
space $\H_{\pi}$ together with a strongly continuous representation
$U_{\pi}$ of the group G such that\\[2mm]
 1) $\pi^{\O_1}  \Y _{\A( \O_2)} = \pi^{\O_2} \ \ \mbox{ if }\ \ \
\O_2 \subset  \O_1 \ \ .$ \\[2mm]
 2) $\mbox{ \it Ad}_{U_{\pi}(g) } \circ \pi^{\O} =
\pi^{g\O} \circ \a_g  \Y _{\A (\O)}\ \ ,$ \\[2mm]
where {\it Ad\/}$_{U_{\pi}(g)}$ is the adjoint action.
For the defining representation on the vacuum sector
$\H^0$ we will use the symbol $\pi_0$.
One often prefers to work with representations of one $C^*$-
algebra instead of representations of the family $\A(\O)_{\O \in \K}$.
This is possible since
every representation of $\A(\O)_{\O \in \K}$ defines a
representation of the $C^*$--inductive limit
\be \A = \overline{\cup_{\O\in \K}  \A(\O)} \ \ .\label{A}\ee
Here the bar denotes closure with respect to the operator norm.
The {\em algebra $\A$ of quasi-local observables} contains all the
local algebras $\A(\O)$.
To perform the inductive limit it is essential that
the sets $\O \in \K$ in the Minkowski space form a directed set.

We will be concerned only with a small subset of representations
which have been singled out because of their relevance
for elementary particle physics \cite{Bor1,Bor2}.
A representation $\pi$ is said to be locally normal
if
\be \pi \Y _{\A(\O)} \cong \pi_0 \Y _{\A(\O)} \ \ \ \mbox{ for all }
\ \ \ \O \in \K\ \ .     \label{lnorm}
\ee
This has a direct physical interpretation. Elements
in $\A(\O)$ describe measurements which can be
performed in $\O$. Typically, different superselection sectors
can be distinguished only by their global properties
(``total charge''). In other words, representations $\pi$
of the algebra $\A$ of quasi-local observables become
equivalent when they are restricted to the local subalgebras
$\A(\O)$. Local normality (\ref{lnorm}) will be tacitly assumed
throughout this text. By definition, the Hilbert space $\H_{\pi}$
carries a strongly continuous representation $\U_{\pi}$ of the
space time translations. When the spectrum of the corresponding
generators is contained in the closed forward light
cone, $\pi$ is a {\em positive energy representation}.

In \cite{DHR1} Doplicher, Haag and Roberts introduced
yet another criterion which selects in general an
interesting subset of positive energy representations.
They called $\pi$
{\em  locally generated with respect to $\pi_0$}, if
\be \pi \Y _{\A (\O')} \cong \pi_0 \Y _{\A(\O')}\ \ .   \label{DHR} \ee
This criterion looks similar to local normality (\ref{lnorm}).
However it is much stronger, since sectors which satisfy
(\ref{DHR}) cannot be distinguished as long as measurements
inside a given region $\O$ are forbidden. One might take
quantum electro dynamics as a counterexample for this
situation, because by Gauss law the charge inside $\O$
can be calculated from he flux through the surface of
$\O$. To measure the latter, one need not  enter the
region $\O$. Since  criterion (\ref{DHR}) is obviously
too restrictive one had to look for generalizations.
Under some additional assumptions, Buchholz and Fredenhagen
have been able to establish a similar criterion (localization in
spacelike cones (``strings''))
which allows to consider all positive energy
representations with an isolated mass shell
in $d \geq 3$--dimensional quantum field theories \cite{BuFr}.
To exploit this
stringlike localization, Haag duality for
spacelike cones $\C$ instead of double cones $\O$ should
be supposed.

Conformal quantum field theories live on the tube $\tilde M_d =
{\bf S}^1 \times {\bf R}$ \cite{LuMa,Mac}. This causes
minor changes in the standard framework described so far.
For our purposes it will suffice to give some details
concerning two--dimensional
conformal quantum field theories.
We parametrize the space
time manifold $\tilde M_2$ by
$(\tau ,\sigma )$ , $\tau=-\infty \dots \infty ,\ \sigma
= 0 \dots 2\pi$. $\tilde M_2$ contains
Minkowski spaces $M_{\zeta}$ as subspaces (see figure 1).\\
\hspace*{1.5cm} \it \begin{picture}(300,180)(00,00)
\setlength{\unitlength}{1.5pt}
\put(00,00){\line(0,1){110}}
\put(60,00){\line(0,1){110}}
\thicklines
\put(00,55){\line(1,-1){45}}
\put(00,25){\line(1,1){45}}
\put(60,55){\line(-1,1){15}}
\put(60,25){\line(-1,-1){15}}
\put(42,37){$M_{\zeta}$}
\put(13,32){$\zeta$}
\put(-2,-10){$0$}
\put(57,-10){$2\pi$}
\put(20,-8){\vector(1,0){20}}
\put(28,-15){$\sigma$}
\put(-10,35){\vector(0,1){20}}
\put(-12,27){$\tau$}
\put(5,10){${\scriptstyle neg.\ timelike}$}
\put(5,5){${\scriptstyle  relative\  to\  \zeta}$}
\put(12,85){${\scriptstyle pos.\ timelike}$}
\put(12,80){${\scriptstyle relative\  to\  \zeta}$}
\put(5,100){$\tilde M_2$}
\put(95,0){
\parbox[b]{2in}{\rm \small
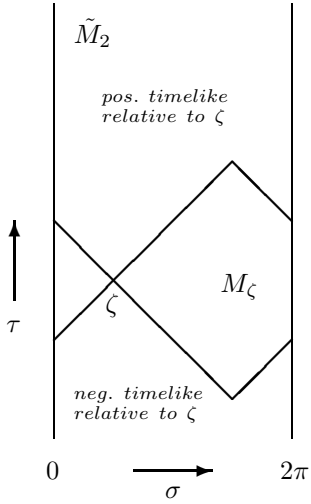
{\bf Figure 1:} M\"obius invariant global causal
structure on the tube $\tilde M_2 = {\bf R} \times {\bf S}^1$.
$M_{\zeta}$ is the Minkowski space  with point $\zeta$
at ``spacelike infinity''. It consists of all points of
$\tilde M_2$ which are relatively spacelike to $\zeta$.}}
\end{picture}\rm \\[15mm]
Their positions are fixed by the unique point $\zeta \in
\tilde M_2$ at spacelike infinity of $M_2$. Manifold $\tilde M_2$
inherits from $M_2$ a global causal structure --- i.e. a
notion of positive timelike, spacelike, and negative
timelike --- which is invariant under the action of
a covering of a direct product of two M\"obius groups on
the light cone coordinates $\s_{\pm}=\frac12 (\tau \mp \s)$. As usual,
the set $\K$ should contain all double cones $\O$ in $\tilde M_2$.
Double cones $\O \in \K$ have the obvious property that their
causal complement $\O'$ in again in $\K$. This implies
the selection criterion (\ref{DHR}) becomes
equivalent to local normality (\ref{lnorm}).
Consequently, {\em every (locally normal) representation
in conformal quantum field theory is locally generated
with respect to the vacuum representations $\pi_0$ \/} \cite{BMT1}.

In two--dimensional conformal quantum field theory
observable fields often  split into mutually commuting
chiral components $\phi_{\pm} (\s_{\pm})$
which depend only on one light cone coordinate.
For such chiral observables, periodicity in
the coordinate $\s$ which parametrizes the
space ${\bf S}^1$ implies periodicity
in time $\t$. This means that $\phi_{\pm}$ can be regarded
as one-valued field on the circle $ \Y z_{\pm} \Y  =1 $, where
$z_{\pm} = e^{i\s_{\pm}}$.
We restrict attention to one chiral component of the
observables and drop the suffix + or --. Since spacelike separated
double cones $\O \in \tilde M_2$ project onto disjoint intervals
$I$ on the circle $ \Y z \Y =1$, {\em chiral observables $\phi(z_1)$
and $\phi(z_2)$ commute when $z_1 \neq z_2, (z_i \in {\bf S}^1)$}.
A set of chiral observables $\phi (z)$ generates a family of
von Neumann algebras $\A(I)$ indexed by intervals $I \in {\bf S^1}$.
To formulate Haag Kastler axioms for such families, double cones
$\O$ should be replaced by intervals $I$. The complement
$I'={\bf S}^1 \setminus I$ substitutes for the causal complement
$\O'$ and one of the M\"obius groups
$SL(2,{\bf R})/{\bf Z}_2$ acts as symmetry group $G$
on the circle ${\bf S}^1$. In a positive energy representation
$\pi$ the generator $L_0$ of ``rotations''  has positive
spectrum. Recently, Brunetti et al. and Fr\"ohlich et al. proved Haag
duality for chiral conformal quantum field theories
\cite{BGL,FrGa2}.

The construction of a $C^*$--algebra of ``quasi
local observables'' is not straight forward since intervals
$I$ in the circle ${\bf S}^1$ do not form a directed
set. To define an inductive limit of local algebras,
one has to remove a point
$\zeta \in {\bf S}^1$
(``point at infinity'').
\be \A_{\zeta} = \overline{\cup_{\I \not\ni \zeta} \A(\I)}
\label{Azeta}\ \ . \ee
Note that after the choice of $\zeta \in {\bf S}^1$,
the complement of an interval $I \not \ni \zeta$ in
${\bf S}^1\setminus \zeta$ decomposes into left and right
components $I_{\<>}$. Disjoint intervals $I_1,I_2 \subset {\bf S}^1
\setminus \zeta$ can be ordered like double cones in
two--dimensional Minkowski space $M_2$.

Even though $\A_{\zeta}$ will suffice for all model independent
studies, it is often inconvenient in the applications.
It is an obvious disadvantage of $\A_{\zeta}$
that local algebras $\A(I)$ are not embedded into
$\A_{\zeta}$ if $\zeta \in I$.
\def\univ{{\mbox{\it \scriptsize univ\/}}}
                               This motivates
to look for a $C^*$--algebra $\A_\univ $ such that
\begin{enumerate}
\item
every local algebra $\A (I)$ can be
embedded into $\A_\univ $ by a unital map $i^{I}$ such
that
$$ i^{I_1} \Y _{\A( I_2)} = i^{I_2} \ \ \ \mbox{ for all }
\ \  I_2 \subset I_1 \ \ $$
and $\A_\univ$ is generated by the algebras
       $i^I(\A(I)),I \subset {\bf S}^1$.
\item
for every representation $\{ \pi^{I}\}_{I \subset {\bf S}^1}$
of the family $\A(I)_{I\subset {\bf S}^1}$
there is a unique representation  $\pi$ of
$\A_\univ $ which satisfies $\pi\circ i^{I} = \pi^{I}$.
\end{enumerate}
The ``universal algebra'' $\A_\univ$ does exist and is unique
\cite{Fre3,FRS2},
but its explicit construction
is subtle. Unlike the algebras $\A$ resp. $\A_{\zeta}$ obtained
from the inductive limits (\ref{A} resp. \ref{Azeta}), the
center of the universal algebra $\A_\univ$ is in general non-trivial.
This means that the vacuum representation $\pi_0$ of
$\A_\univ$ may not be faithful. Indeed it is possible that
two charge operators localized in domains $I_1,I_2, I_1\cup I_2
={\bf S}^1$ add up to to a global quantity which commutes with
all elements in $\A_\univ$. These charges may have different values
in different superselection sectors. They must not be
identified with multiples of the identity.

The setup for theories with charges localized along strings
in 3--dimensional Minkowski space
is very similar to the
situation in chiral conformal quantum field theories
(cf. \cite{FRS2} and references therein).
Points $\zeta $ on the circle ${\bf S}^1$  are substituted
by directions in the two--dimensional
space and one uses spacelike cones $\C$
instead of intervals $I \subset {\bf S}^1$.

\subsection{Localized endomorphisms and fusion structure}

A detailed analysis of the structure of superselection
sectors was first performed by Doplicher,
Haag and Roberts \cite{DHR1,DHR2}.
It was restricted to sectors which are locally
generated with respect to the vacuum sector and formulated
for theories on the four--dimensional Minkowski space.
The generalization to string-like localized sectors
\cite{BuFr} gives essentially the same structure.
If the dimension of the space time manifold is decreased,
specific new features appear. In d=2 (resp. d=3) dimensional
space-times, double cones (resp. spacelike cones) can be
ordered and -- as we will explain below --
this gives rise to representations of the braid
group. Such situations were considered more recently by
Fredenhagen, Rehren and Schroer \cite{FRS1,FRS2}.
Our short exposition will concentrate on representations
localized in double cones $\O$ in
Minkowski space. We included some remarks relevant to treat
chiral conformal  theories and string-like localized sectors
in three--dimensional quantum field theory.
For many further results
and discussions the reader is referred to the original
papers -- especially to \cite{FRS1,FRS2} -- and the reviews
of Fredenhagen \cite{Fre2} or by Kastler, Mebkhout and
Rehren (in \cite{KasB}). Theories with charges localized
in spacelike cones in $M_3$ have been considered
explicitly by Fr\"ohlich et al. \cite{FGM,FrMa,FrGa1}.
Applications to models in two-dimensional conformal
quantum field theory can be found in \cite{BMT1,MSI,MSII,FGV}.

{}From now on, families $\A(\O)_{\O\in \K}$ are always
supposed to satisfy Haag-Kastler axioms and Haag duality (\ref{Hdual}).
Except from some remarks, representations are assumed to be
localized on double cones (\ref{DHR})
$\O$ in Minkowski space $M_d$. Equivalence classes
of locally generated positive energy representations form a set
denoted by $\Rep$.

The notion of {\em localized endomorphisms} will provide the
key to all further analysis. By definition, an endomorphism
$\rho$ of the $C^*$ algebra $\A$ is a linear map
$\rho : \A \to \A$ with the properties \vspace*{-4mm}
\bea \rho (AB) & = & \rho (A) \rho (B) \ \ ,\nn \\
\rho (A^*) &=& \rho (A)^* \ \ ,\nn\\
\rho ({\bf 1}) &=&  {\bf 1}\ \ . \nn
\eea
It is called an {\em automorphism} if it has an inverse. Endomorphisms
of $\A$ fall into equivalence classes $[\rho]$ with respect
to inner automorphisms, i.e. conjugation by unitaries $U \in \A$.
$\rho$ is said to be {\em localized} in $\O \in \K$ if
$$ \rho ( A ) = A \ \ \ \mbox{ for all } \ \ \
A \in \A(\O_1)\subset \A \ ,
\ \ \O_1 \subset \O' \ \ .$$ An endomorphism $\rho$
localized in $\O$ is called {\em transportable} whenever
equivalent morphisms $\s \in [\rho]$ localized in the transformed
region $g\O$ exist for all $g \in G$. Transportable
endomorphisms localized in spacelike separated regions commute
\cite{DHR1}.

Endomorphisms of $\A$ can be used to obtain positive
energy representations $\pi_0 \circ \rho$ of $\A$ on
the Hilbert space $\H^0$. For every locally generated
positive energy representation $\pi$ of $\A$ there is a
localized transportable endomorphism $\rho$ such that
\be \pi \cong \pi_0 \circ \rho \ \ \ .\label{ppcr}\ee
This can be seen as follows. According to the criterion (\ref{DHR}),
$\pi \Y _{\A(\O')}$ is unitary equivalent to $\pi_0 \Y _{\A(\O')}$,
i.e. for each double cone $\O$ there is a unitary
$V: \H_{\pi} \to \H^0 $ such that
$$ V \pi (A) = \pi_0 (A) V  \ \ \ \mbox{ for all } A \in \A(\O')\ \ . $$
By Haag duality, the map
$$ \rho (A) = V \pi (A) V^*$$
defines an endomorphism of $\A$ localized in $\O$.
It is transportable and has the desired property (\ref{ppcr}).
Two localized endomorphisms $\rho_i,i=1,2$ are equivalent
if and only if the representations $\pi_0 \circ \rho_i$
are equivalent. We conclude that elements of $\Rep$, i.e. equivalence
classes of locally generated positive energy representations
of $\A$, correspond one by one to equivalence classes $[\rho]$
of localized transportable endomorphisms $\rho$. This observation
will be used to identify both objects.

Let us pause for a moment to comment on (chiral) conformal
quantum field theory. In this case
superselection sectors carry irreducible
positive energy representations
of the $C^*$--algebra $\A_{\zeta}$ introduced in the last
section. By the results in \cite{BMT1,BGL,FrGa2}, every
positive energy representation is obtained as a
composition $\pi_0 \circ \rho$ of the vacuum
representation $\pi_0$ with a
endomorphism $\rho$ of $\A_{\zeta}$. We can assume $\rho$ to be
localized and transportable in an appropriate sense.
The general considerations below can be established on
$\A_{\zeta}$ without modifications.
In practice it will be more convenient to work with
endomorphisms $\rho$ of the
universal algebra $\A_\univ$. Fredenhagen has shown \cite{Fre3} that
the localized and locally transportable endomorphisms of $\A_\univ$
exist for all elements of $\Rep$. They
restrict to localized transportable endomorphisms of
$\A_{\zeta} \subset \A_\univ$ when $\zeta$ lies outside
the domain of localization. Basically,  these
remarks apply also to stringlike localized sectors in
2+1 dimensional quantum field theory.

All this is much more than a technicality. Endomorphisms of
$\A$ can be composed and thus lead to a proper definition
of a product of sectors. Given two representations $\pi_i = \pi_0
\circ \rho_i, i=1,2$ on the Hilbert space $\H^0$, their
product $\pi_1 \ti \pi_2$ is defined by
$$\pi_1 \ti \pi_2 =
\pi_0 \circ \rho_1 \circ \rho_2 \ \ \ . $$
The equivalence class of
$\pi_1 \ti \pi_2$ is an element of $\Rep$ if
equivalence classes of
$\pi_i,i=1,2,$ are. In other words a product in $\Rep$ is
defined by $[\rho_1] \ti [\rho_2] \equiv [\rho_1 \circ \rho_2]$.
These
assertions do not depend on the type of localization.

Two representations
$\pi_0 \circ \rho_i, i=1,2,$ in $\H^0$ have
equivalent subrepresentations if
an isometry  $U \in \B(\H^0)$ intertwines between them, i.e.
$U \pi_0(\rho_1(A)) =  \pi_0(\rho_2(A))U$ for all
local observables $A \in \A $. When restricted with the
''source'' projection  $U^*U$ of $U$,
the representation $\pi_0 \circ \rho_1$
is equivalent to the restriction of $\pi_0 \circ \rho_2$ to the
range of $U$. By Haag duality,
such intertwining operators $U$ can be obtained
as image of a local intertwiner $T \in \A$
$$ T \rho_1 (A) = \rho_2 (A) T \ \ \ \mbox{ for all} \ \ A \in \A\ \ .$$
Local observables $T \in \A$ which satisfy this equation
span a complex linear space  $\T(\rho_1,\rho_2)$. If $S,T \in
\T(\rho_1,\rho_2)$ then $TS^*$ is a local observable in $\T(\rho_2,
\rho_2)$. Schurs lemma asserts, that irreducibility of $\rho$
implies $\T(\rho,\rho) \cong {\bf C}$. In conclusion, $<T,S> \equiv
TS^*$ defines a scalar product on $\T(\rho_1,\rho_2)$ if $\rho_2$
is irreducible. Therefore $\T(\rho_1,\rho_2)$ is a Hilbert space in
this case.

The product of sectors $[\rho_i],i=1,2,$ is commutative in the
sense that $[\rho_1 \circ \rho_2 ] = [\rho_2 \circ \rho_1]$.
To see this we pick two endomorphisms $\s_i$ from the equivalence
classes $[\rho_i]$  which are localized in spacelike
separated double cones. As we remarked above, their action
on the observables commutes and so the assertion follows.
For every pair of localized and transportable endomorphisms $\rho_i,i=1,2$
there is a unitary local intertwiner $\e(\rho_1,\rho_2) \in
\T(\rho_1\circ \rho_2, \rho_2 \circ \rho_1)$, the {\em statistics
operator}. The collection of statistics operators
in uniquely determined by the following
equations (a detailed proof can be found in the contribution
of Mebkhout et al. in \cite{KasB}).
\bea
\e(\rho_1,\rho_2) \rho_1 (T_2 ) T_1 & = & T_2 \s_2 (T_1) \e(\s_1,\s_2)
\ \ \ \ \mbox{ for all } \ \ \ T_i \in \T(\s_i,\rho_i) \nn \ \ ,\\[2mm]
\e(\rho_1 \circ \rho_2, \s ) = \e(\rho_1,\s ) \rho_1(\e (\rho_2,\s ))
\ \ \ &  & \ \ \ \e(\s,\rho_1 \circ \rho_2) = \rho_1 (\e (\s,\rho_2))
\e(\s, \rho_1) \label{epsprop}\ \ ,\\[2mm]
\e (\rho_1,\rho_2) &=& 1 \ \ \ \mbox{ whenever } \ \ \ \rho_1  > \rho_2
\ \ .\nn \eea
In the last row, $\rho_1 > \rho_2$ refers to the order of
localization regions, provided they can be ordered.
For localization in double cones contained in a two--dimensional
Minkowski space this is the case and $\rho_1 > \rho_2$
means that $\rho_1$ is localized on a domain
$\O_1$ left from the localization
region $\O_2$ of $\rho_2$, i.e. $\O_1 > \O_2$.
Trivialization for $\rho_1< \rho_2$ would give rise to the
opposite statistics operator $\e(\rho_2,\rho_1)^*$. In higher
dimensional quantum field theories there is no invariant
distinction between left and right so that trivialization
is possible for all pairs of spacelike separated endomorphisms.
For two--dimensional light cone theories [stringlike localized
sectors in theories on $M_3$], the notion $\<>$ refers to
the point $\zeta$ at infinity [the direction in the two--dimensional
space] which we have to single out to define the inductive
limit in (\ref{Azeta}).

It follows from these relations that the vacuum sector carries
a representation of the colored braid group. In particular,
$\s_i = \rho^{i-1}(\e(\rho,\rho))$ satisfy Artin
relations (\cite{Art}).
\bea       \label{Artinu}
\sigma_i \sigma_k = \sigma_k\sigma_i \ \ \mbox{ if }
 \Y k-i \Y  \geq 2   & , & \sigma_i \sigma_{i+1} \sigma_i =
\sigma_{i+1} \sigma_i \sigma_{i+1} \ , \\[1mm]
\sigma_i \sigma_i^{-1} = & {\bf 1} & = \sigma^{-1}_i \sigma_i \ \ ,
\label{Artinl} \eea
i.e. the elements $\s_i$ and $\s_i^{-1}$ $(i=1
\dots n-1)$ generate the braid group $B_n$. In theories
without an invariant distinction between left and right,
$\s_i =\s_i^{-1}$ so that we obtain a representation of
the permutation group. Accordingly the statistics
operators give rise to an intrinsic notion of
statistics of a superselection sector.

At this stage we would like to restrict to sectors of
{\em finite statistics}. An irreducible transportable
endomorphism has finite statistics whenever it
possesses a left inverse \footnote{By definition, a
left inverse is a positive linear map $\Phi: \A \mapsto \A$
such that $\Phi(\rho(A) B \rho(C)) = A \Phi(B) C$ for
all $A,B,C \in \A$ and $\Phi({\bf 1}) = {\bf 1}$.}
$\Phi$ with $\Phi(\e(\rho,\rho))
\not = 0$. It has been shown by Longo \cite{Lon}
that this is equivalent to a finite Jones index
\cite{Jon} of the inclusion $\rho(\A (\O)) \subset \A(\O)$.
We do not plan to justify this restriction (see however
\cite{Fre1,BuFr}) but list the consequences
required later. Whenever two sectors of finite
statistics are composed, their product decomposes
into a finite direct sum of sectors which have finite statistics
again. This means that the subset
$\Rep' \subset \Rep$ of finite
direct sums of sectors with
finite statistics is closed
under products.
Moreover, every sector with finite statistics
has a unique conjugate, i.e. for every irreducible
endomorphism $\rho$ of $\A$ there exists a unique
irreducible endomorphism $\bar \rho$ such that
$\bar \rho \circ \rho$ contains the vacuum sector.
One can show that $dim \T(\bar \rho \circ \rho,id) =1$.

Equivalence classes of irreducible representations
in $\Rep'$ will be labelled by elements $I,J,K,\dots $
of an index set $\I_{\A}$. We reserve $0\in \I_{\A}$ for the
vacuum sector. Let us fix a set of representative
endomorphisms $\rho_I$, one for each superselection sector
in $\Rep'$. According to the general remarks on sectors
with finite statistics, the product of two endomorphisms
$\rho_I \circ \rho_J$ is equivalent to a finite
direct  sum of irreducibles $\rho_K$. As we saw above,
the space $\T(\rho_I \circ \rho_J  \Y  \rho_K)\equiv \T(IJ \Y K)$
is a Hilbert space. The dimension $dim(\T(IJ \Y K))$ defines
the fusion rules $N^{IJ}_K$ which appear in the decomposition
$$ [\rho_I \circ \rho_J] = \bigoplus_K [\rho_K] N^{IJ}_K \ \ .$$
It follows from the associativity of the composition of
endomorphisms and the commutativity we found above that
the fusion rules are associative and commutative in the
sense
\bea
N_K^{IJ} &=& N_K^{JI} \ \ \ \mbox{ and } \\
\sum_M N_L^{IM} N_M^{JK} &=& \sum_M N_L^{MK} N_M^{IJ}\ \ .
\label{comass} \eea
The endomorphism $\rho_{\bar J}$ should be the unique
conjugate $\bar \rho_J $ of $\rho_J$. Since the vacuum
sector $[\rho_0]$ appears with multiplicity 1 in
the decomposition of $[\rho_{J} \circ \rho_{ \bar J}]$
we have
$$ N^{I \bar J}_0 = \delta_{I,J}\ \ .  $$

In the Hilbert space $\T(IJ \Y K)$ we choose
an orthonormal basis $T_a\vvert{K}{I}{J} \in \A$.
In detail this means that the operators $T_a \vvert{K}{I}{J},
a = 1  \dots N^{IJ}_K,$
satisfy
\bea   \label{Tint}
T_a \vvert{K}{I}{J} \rho_I \circ \rho_J (A) & = &
\rho_K(A)T_a\vvert{K}{I}{J} \ \ \ \mbox{ for all } A \in \A \ \ ,\\[2mm]
T_a \vvert{K}{I}{J} T_b \vvert{L}{I}{J}^* = \delta_{a,b} \delta_{L,K}
& , & \sum_{K,a} T_a \vvert{K}{I}{J}^* T_a\vvert{K}{I}{J} =1\ \ .
\eea
The {\em fusion- and braiding- matrices} are complex valued matrices
defined by
\bea
T_a \vvert{L}{M}{K} T_b \vvert{M}{I}{J}
& = & \sum_N T_c \vvert{L}{I}{N} \rho_I (T_d \vvert{N}{J}{K})
 \Fus{M}{N}{K}{J}{L}{I}^{cd}_{ab} \ \ ,\label{Fusdef} \\
T_a \vvert{L}{M}{J} T_b \vvert{M}{I}{K}  \rho_I
(\e (\rho_J,\rho_K))
& = & \sum_N T_c \vvert{L}{N}{K} T_d \vvert{N}{I}{J}
\Br{M}{N}{J}{K}{L}{I}^{cd}_{ab}   \ \ .  \label{Brdef}
\eea
As a special case of eq. (\ref{Brdef}) for $I=0$ we introduce
the matrix $\Omega\vvert{L}{M}{J}$.
\be  \label{Omegadef}
T_a\vvert{L}{K}{J} \e (\rho_J,\rho_K) \equiv T_b \vvert{L}{J}{K}
\Omega^{b}_{a} \vvert{L}{K}{J}  \ \ .
\ee
By Schurs' lemma, the coefficients $F,B,\Omega$ are
certain complex matrices. They are determined by the
model and depend only
on the equivalence classes $[\rho_I]$ but not on $\rho_I$ itself.
The braiding matrices $B , \Omega$ are often denoted by
$\Omega(+), B(+)$ to emphasize their dependence on the choice
of trivialization of the statistics operators $\e(\rho_J,\rho_K)$
(\ref{epsprop}). The corresponding matrices for the opposite statistics
operators $\e(\rho_K, \rho_J)^*$ are called
$B(-), \Omega (-)$. We restrict attention to one
trivialization and neglect to write $(\pm)$.
A short  calculation reveals  the
following proposition.

\begin{prop}{\em (Polynomial equations) \cite{FRS1}}\label{poly}
The fusion- and braiding matrices  $F,\Omega$
defined by (\ref{Brdef},\ref{Fusdef})
solve the polynomial equations
\bea
\sum_N \Fus{N}{P}{K}{J}{L}{I} (\Oe{L}{K}{N} \o 1 )
\Fus{M}{N}{J}{I}{L}{K} & = &
(1 \o \Oe{P}{K}{J} ) \Fus{M}{P}{J}{K}{L}{I}
(1 \o \Oe{M}{K}{I} ) \label{Hex}\\
\sum_Q \Fus{Q}{S}{L}{K}{R}{J}_{23} \Fus{N}{R}{L}{Q}{P}{I}_{12}
\Fus{M}{Q}{K}{J}{N}{I} _{23} & = &
P_{23} \Fus{M}{R}{S}{J}{P}{I}_{13}
\Fus{N}{S}{L}{K}{P}{M}_{12}\label{Pent}\\
\sum_P \Fus{P}{Q} {I}{J}{K}{L} \
\Fus{P}{R}{I}{J}{K}{L}^* ={\bf 1} \d_{Q,R} &  &
\sum_Q \Fus{P}{Q} {I}{J}{K}{L}^* \
\Fus{R}{Q}{I}{J}{K}{L} ={\bf 1} \d_{P,R}
\label{norm}\\[1mm]
\Oe{L}{J}{I} \Oe{L}{J}{I}^* = {\bf 1} & &
\Oe{L}{J}{I}^* \Oe{L}{J}{I} = {\bf 1} \\[3mm]
\Fus{I}{J}{J}{0}{K}{I} &=&  {\bf 1} 
\label{trivial}
\eea
The braiding-matrix $B$ can be calculated from
$F,\Omega$.
\be
\Br{M}{N}{J}{K}{L}{I} = \sum_P \Fus{N}{P}{K}{J}{L}{I}^*
(1 \o \Oe{P}{K}{J}) \Fus{M}{P}{J}{K}{L}{I} \ \ .
\label{Brmat} \ee
We used an obvious matrix notation and ${\bf 1}$
denotes an appropriate unit matrix. $F_{12}$ is defined
on threefold tensor products $ u \o v \o w $ by
$ F_{12} (u \o v \o w) = F (u\o v) \o w $ etc.
$P_{23}$ acts as permutation of the second and third
component.
\end{prop}

The first two relations (\ref{Hex},\ref{Pent})
are the famous Moore-Seiberg ``hexagon'' and
``pentagon'' identities known from conformal
quantum field theory
\cite{MoSe1,MoSe2}.
In our context they reflect deep properties
of the fusion structure of superselection
sectors. In particular, conformal symmetry
was not assumed.

\noindent
{\sc Proof:} We do not want to prove all the relations
but just demonstrate the type of calculations to be
done at the example of eq. (\ref{Brmat}).
The product of operators which appears on the left hand side
of eq. (\ref{Brdef}) can be manipulated in two
different ways. One is just the step from the
left to the right hand side of (\ref{Brdef}). For the
other we apply the definition (\ref{Fusdef}) of the
fusion matrix, the endomorphism property of $\rho_I$,
the definition (\ref{Omegadef}) and relation (\ref{Fusdef})
in this order.
\bea
T_a \vvert{L}{M}{J} T_b \vvert{M}{I}{K}  \rho_I
(\e (\rho_J,\rho_K)) & = &
 \sum_P T_c \vvert{L}{I}{P} \rho_I (T_d \vvert{P}{K}{J}) \rho_I
(\e (\rho_J,\rho_K)) \Fus{M}{P}{J}{K}{L}{I}^{cd}_{ab}  \nn \\
  &= &
 \sum_P T_c \vvert{L}{I}{P} \rho_I (T_e \vvert{P}{J}{K})
 \Oe{P}{K}{J}^e_d \Fus{M}{P}{J}{K}{L}{I}^{cd}_{ab}  \nn \\
   &= &
 \sum_{NP} T_f \vvert{L}{N}{K} T_g \vvert{N}{I}{J}
 (\Fus{N}{P}{K}{J}{L}{I}^*)^{fg}_{ce}
 \Oe{P}{K}{J}^e_d \Fus{M}{P}{J}{K}{L}{I}^{cd}_{ab}\nn .
\eea
We used the first relation in (\ref{norm}) for the last
equality. If we compare the result with the right hand
side of (\ref{Brdef}) we find that the same operator
has been expressed by two  linear combinations of the
same basis elements. Consequently, the coefficients
have to agree and this gives eq. (\ref{Brmat}).

\subsection{Covariant field operators}

Now we are prepared to construct field operators $\Psi$
which make transitions between different superselection sectors.
We want them to transform non-trivially under the action
of elements $\xi$ from an ``appropriate'' symmetry algebra $\G^*$.
A simple assumption on the structure of the representation theory
of $\G^*$ will turn out to ensure
the existence of such covariant field operators.
They will be constructed as a sum of ``vertex operators''.

$\G^*$ is assumed to be
a semisimple bi-*-algebra with antipode.
                             Equivalence classes $ [\t]$
of finite dimensional irreducible representations $\t$ of $\G^*$ are
labelled by elements of an index set $\I_{\G^*}$.
Because of semisimplicity \footnote{Note that every finite
dimensional representation
of a semisimple algebra is a
direct sum of irreducible representations.},
tensor products of
two finite dimensional irreducible representations $\t,\t'$ on vector
spaces $V,V'$ can be decomposed into irreducibles
$\t^{\alpha} $ on $V^{\alpha}$ . The corresponding ``Clebsch
Gordon'' intertwiners $C(\t \bo \t' \Y \t^{\alpha}) : V \o V'
\mapsto V^{\alpha}$ form complex vector spaces
$\C(\t \bo \t' \Y  \t^{\alpha})$.
$$    C(\t \bo \t'  \Y  \t^{\alpha} ) (\t \bo \t')(\xi)
= \t^{\alpha} (\xi) C(\t \bo \t'  \Y  \t^{\alpha} ) \ \ .$$

Recall that a representation $\t$ of $\G^*$ on a Hilbert space
$V$ is unitary, if $\t(\xi)^* = \tau(\xi^*)$ for all $\xi \in \G^*$\\
{\bf Assumption:} \it Let $I_{\A}$ denote the set of
superselection sectors with finite statistics as in the
preceding subsection. We assume that there is a bijection
$\theta:\I_{\A} \mapsto \I_{\G^*}$ and a set of
unitary representatives $\t^{\theta(I)}$ from the
equivalence classes of irreducible representations
of $\G^*$ such that
\be \mbox{ dim\/}(\C(\t^{\theta (I)}\bo \t^{\theta (J)} \Y
\t^{\theta (K)})) = N^{IJ}_K \ \ . \label{fundAss} \ee
\rm

In other words: There is a unique equivalence class of irreducible
representations of $\G^*$ associated with every superselection sector
and fusion rules of the sectors are
in agreement with the selection rules of the prospective
symmetry. By uniqueness of the conjugate,
$\theta$ will automatically map the vacuum sector
to the equivalence class of the one--dimensional
trivial representation $\e$ of $\G^*$. Since $\e$ is unitary
we can always choose
$$\t^{\theta (0)} = \e\ \ .$$
We will not distinguish between $I$ and $\theta(I)$ in the
following.

Given an algebra of observables $\A$ one may wonder whether
a bi-*-algebra with antipode satisfying the above assumption does exist.
We postpone this discussion until the next section.
It will turn out that
{\em suitable bi-*-algebras with antipode
do always exist}, if $\I_{G^*}$ is finite (cp. corollary
\ref{Ex} below).

Given the required  bi-*-algebra with antipode
$\G^*$, we can start to
build covariant field operators. They will act on a
Hilbert space $\H$ of physical states which is a
direct sum of irreducible representation spaces
$\H^J$ for the algebra of observables $\A$, each with
multiplicity $\delta_J$ determined by the dimension of
the representation $\t^J$ of $\G^*$, i.e.
$ \delta _J \equiv dim ( \tau^J)$.
\be
\H = \bigoplus_{J \in \I } \bigoplus^{\delta_J}_1 \H^J_i
\ . \ee
$\H^0$ carries the vacuum representation $\pi_0$ of $\A$ and it
occurs with multiplicity one (since $dim(\tau^0) = dim(\e ) = 1$).

Next we define a representation $\pi$ of
the observables algebra $\A$ on all of $\H $ by
its restrictions to the subspaces $\H^J_m$,
\be \pi (A) =\pi_J (A)\ \mbox{on }\ \H^J_m  \ (m=1...\delta_J) \ . \ee
According to our general discussion, the representation $\pi_J$
on $\H^J_m$ is equivalent to $\pi_0 \circ \rho_J$ which is realized
on the vacuum Hilbert space $\H^0$. This equivalence can be expressed
by isometries $i_{Jm}^{\ast }: \H^0 \mapsto \H^{J}_m$ with
the intertwining property
\be  \pi_J(A)i_{Jm}^{\ast } =
             i_{Jm}^{\ast } \pi_0(\rho_J(A))  \ . \label{iint}\ee

To specify the action of elements $\xi \in \G^*$ on
$\H$, we choose an orthonormal basis $e^J_m$ in the
finite dimensional representation space $V^J$. When
the corresponding matrix elements of $\t^J(\xi)$ are
denoted by $\t^J_{km}(\xi)$, an unitary
representation $\U$ of $\G^*$ on $\H$ is obtained
according to
\be \U(\xi )i^{\ast }_{Jm} \Y \psi >=
            i^{\ast }_{Jk} \Y \psi >\tau^J_{km}(\xi ) \label{Ustate}\ee
for arbitrary $ \Y \psi > \in \H^0 $ .
The symmetry acts as a \it gauge symmetry
(of first kind), \rm i.e. all
observables are invariant
\be  [\U(\xi ),\pi( A)]=0 \ \ \mbox{for all}
\ A\in \A, \ \xi \in \G^* \ . \ee

Our main task is to construct field operators $\Psi^J_m(\rho_J)$
which make transitions between the sectors $\H^I_i$
with different $I$. They will have the following properties
\begin{enumerate}
\item
Intertwining property for representations of $\A $,
\be \pi (A) \Psi^J_m(\rho_J) = \Psi^J_m(\rho_J) \pi (\rho_J(A))\ .  \ee
As a consequence, $\Psi^J_m(\rho_J)$ commutes with observables localized
in the causal complement $\O'$ of the localization region $\O$ of
the endomorphism $\rho_J$. This property reflects locality of the
field operators with respect to observables.
\item
 Field operators $\Psi^J_m(\rho_J)$ transform covariantly according
 to the representation $\t^J$ of $\G^*$.
\be
 \U(\xi )\Psi^J_m(\rho_J) = \Psi^J_k(\rho_J)
 (\tau^J_{km}\o \U )(\D (\xi ))\ \ .
\ee
\end{enumerate}
The field operators $\Psi^J_m(\rho_J)$ are determined by these
properties up to a phase factor. They will be build up
from the following ``vertex operators''.
\bea
\ _k\Psi_l\vvert{K}{L}{J}_{a}\/ (\rho_J) & : & \H^L_l \mapsto \H^K_k \ \
\mbox{ for } \ a= 1 \dots N^{JL}_K\ \ , \\[2mm]
\ _k\Psi_l\vvert{K}{L}{J}_{a}\/ (\rho_J) & = &
   i^*_{Kk}  \pi_0 (T_a \vvert{K}{L}{J}) i_{Ll}\ \ .
\eea
Combining the intertwining relations (\ref{iint})
and (\ref{Tint}) we find that these operators satisfy
the intertwining property {\it 1.} for representations of
$\A$. We extend the vertex operators $\ _k\Psi_l\vvert{K}{L}{J}_a$
to all of $\H$
such that they vanish on states in $\H^I,i$ for all
$i\neq l , I \neq L$. The extended operators are denoted by
the same symbol. Their intertwining properties with
observables is not affected by this extension.

To obtain a covariant field operator $\Psi^J_j(\rho_J)$
on the whole Hilbert space $\H$, we fix an orthonormal
basis $C^a (JL \Y K),a = 1 \dots N^{JL}_K$  in the
Hilbert spaces $\C(JL \Y K)\equiv \C(\t^J \bo \t^L \Y \t^K)$
of Clebsch Gordon intertwiners. By assumption, the fusion rules
and tensor product decomposition match so that the index $a$
assumes the same values as for the vertex operators. With this
knowledge we can form the following linear combination of
vertex operators.
\be
\Psi^J_m(\rho_J) = \sum_{K,k,L,l}
\ _k\Psi_l\vvert{K}{L}{J}_a\/(\rho_J)
 \CG{J}{L}{K}{m}{l}{k}^a
\ \ \   . \label{Fieldop}
\ee
The complex coefficients $\CG{.}{.}{.}{.}{.}{.}^a $ are
matrix elements of the Clebsch Gordon map $C^a(JL \Y K)$
in the basis $e^J_j \o e^L_l$ resp. $e^K_k$. The field
operators $\Psi^J_m(\rho_J)$ meet all the requirements
stated before.

\begin{theo} \label{fieldalg}
Let $\A(\O)_{\O \in \K} \subset \B(\H^0)$ be a family
of local observable algebras with properties as
before. Suppose that $\G^*$ is a bi-*-algebra
which satisfies assumption (\ref{fundAss}).
Then there is a representation $\pi$ of $\A$ on a
Hilbert space $\H$, a unitary representation $\U$ of $\G^*$
on $\H$ and a family $\B(\O)_{\O \in \K} \subset \B(\H)$
of $*$-algebras such that
\begin{enumerate}
\item
the vacuum representation $\pi_0$ is a subrepresentation
of $\pi$ on $\H^0 \subset \H$ and appears with
multiplicity 1. States in $\H^0$ are invariant with
respect to the action of $\G^*$. In particular,
$$ \U(\xi) \vac = \vac \e(\xi ) \ \ \ \mbox{ for all }
\ \ \ \xi \in \G^* \ . $$
The Hilbert space $\H$ is generated from $\vac$ by
algebras $\B(\O) \subset \B(\H)$.
\item
the algebras $\B(\O)\subset \B(\H)$ are generated by
elements of $\A(\O)$  and operators $\U(\xi)$ together with operators
$\Psi^J_m (\rho_J)$, where $\rho_J$ are endomorphisms
of $\A$ localized in $\O$. The field operators
$\Psi^J_m(\rho_J)$ are local relative to
observables and transform covariantly according
to the representation $\t^J$ of $\G^*$. Explicitly
this means
\bea
\Psi^J_m(\rho_J) \pi(A)  & = & \pi(A) \Psi^J_m(\rho_J)
\ \ \ \mbox{ for all }
\ \ A \in \A(O') \ \ , \\[2mm]
\U(\xi )\Psi^J_m(\rho_J) & = &
 \Psi^J_k(\rho_J) (\tau^J_{km}\o \U )(\D (\xi ))\ .
\eea
\item
each equivalence class of irreducible representations in $\Rep'$
(i.e. locally generated sector with finite statistics)
is realized as a subrepresentation of $\pi$.
\end{enumerate}
\end{theo}

For these results the existence of an antipode $\S$ was superfluous.
The antipode is used to define a covariant adjoint of field
operators (proposition \ref{COVADJ}).
Recall that a controlled transformation behaviour of adjoint fields
requires in addition an element $\vp$ satisfying (\ref{vpiint}).
It  will be argued later that the assumptions of the preceeding theorem
suffice to construct such intertwiners $\vp$ without any further input.

We note that operators $\pi(A)$ and $\U(\zeta)$ with $A \in \A(\O)$
and $\zeta$ in the center of $\G^*$ generate an invariant
subalgebra of $\B(\O )$ (i.e. elements commute with $\U(\xi)$
for all $\xi \in \G^*$.
This implies that not all invariants in the algebra $\B(\O)$
are observables of the model. A detailed discussion is
given below.

For chiral conformal quantum field theories, field operators
can be obtained in the same way. It was explained above that the
corresponding
endomorphisms $\rho_J$ act on an algebra $\A_{\zeta}$
which depends on the choice of the ``point at
infinity'' $\zeta \in {\bf S}^1$. This dependence on $\zeta$
does also show up in the field operators $\Psi^{\zeta J}_m(\rho_J)$
and the algebras of fields $\B^{\zeta}(I)$. If $\zeta$ is changed,
field operators are multiplied with a unitary element
from the center of the universal algebra $\A_\univ$.
This means that fields do not ``live'' on the circle
${\bf S}^1$ but on a covering thereof. The same
behaviour is found for field operators which create
charges localized along strings in a three--dimensional
Minkowski space.
For an enlightening discussion of these points the reader is
referred to \cite{FRS2}.

The field operators $\Psi^J_m(\rho_J)$ are localized
in the localization domain of $\rho_J$. One may construct
operators $\Psi^J_m(x,t)$ associated with a point $(x,t)$ by
taking appropriate limits.
If $\rho_J^\a$ is a sequence of endomorphisms from the
equivalence class of $\rho_J$ such that the localization
region shrinks to a point $(x,t)$ in the limit $\a \rightarrow \infty$,
$\Psi^J_m(x,t)$ is obtained formally as
$$ \Psi^J_m(x,t)= \lim_{\a \rightarrow \infty} \N_{\rho_J^\a}
\Psi^J_m (\rho_J^\a) \ \ , $$
where $\N_{\rho_J^\a }$ are suitable normalization factors.
This procedure has been successfully applied to charged
fields of the $U(1)$-current algebra on the circle ${\bf S}^1$
\cite{BMT1} and there is much hope to develop a general
technique for chiral conformal quantum field theories
\cite{Joe,FrJo}.

\section{Weak Quasi Quantum Groups}  \setcounter{equation}{0}
The formulation of quantum
symmetry in section 2 involved only a bi-*-algebra structure.
One cannot expect that every bi-*-algebra is
actually realized as a quantum symmetry of a
quantum mechanical system. At this point it
should suffice to remark that
only group symmetries seem to be
realized in higher dimensional quantum field theory.
We will elucidate the reasons later.

To describe distinguished algebraic structures
within the class of bi-*-algebras we introduce
and discuss some relevant notions.
Weak quasi quantum groups will be defined. They
were introduced in \cite{MSIII} as a generalization
of Drinfeld's quasi quantum groups \cite{Dri2}.
References to our physical framework
have been avoided to emphasize the purely mathematical
nature of the arguments. Our presentation is restricted
to those parts of the theory which are needed later.

\subsection{Fundamental definitions and results}

In a bi--algebra, tensor products of representations
are defined with the help of the co-product $\D:\G^* \mapsto
\G^* \o \G^*$. Properties of the tensor product (\ref{tensprod})
of representations can be traced back to properties of
the co-product.
As an example consider
the group algebra associated with a compact group G.
In this case, the tensor product (\ref{gtensprod}) of
representations is well known to be associative and commutative.
This corresponds to a co-commutative and co-associative
co-product $\D_G$. A co-product $\D$ is called
co-associative, if
\be (\D \o id) \D (\xi) =
(id \o \D) \D (\xi)
\ee
and co-commutativity means that
\be \D(\xi) = \D'(\xi) \ \ .\ee
Given the expansion $\D(\xi ) = \sum \xi^1_{\s} \o \xi^2_{\s}$\ ,
$\D'$ is defined by
$\D'(\xi) = \sum \xi^2_{\s} \o \xi^1_{\s}$.
For $\D_G$
both properties can be verified from the explicit
action (\ref{DeltaG}) of $\D_G$ on elements in G.
In the following we introduce a special class of
bi--algebras for which tensor products of representations
are at least commutative and associative {\em up to
equivalence}.
\begin{defn} {\em (quasi--co-associativity) \cite{Dri2,MSIII}}
 \label{Defquasiass}
The co-product $\D$ of a
bi-*-algebra with antipode $(\G^*,\D,\e,*,\S)$
is called {\em quasi--co-associative},
if an element $\vp \in \G^*\o \G^*
\o \G^*$ exists, such that
\begin{enumerate}
\item
$\vp$ has a quasi-inverse $\vp^{-1}
\in \G^* \o \G^*\o \G^*$
such that
\be
\vp \vp^{-1} =
(id \o \D) \D(e)\ \ \ \  , \ \ \ \ \vp^{-1} \vp = (\D \o id) \D(e)\ \ .
\label{quasiinvvp} \ee
\item
$\vp$ satisfies the ``intertwining'' relation
\be
\vp (\Delta \o id )\Delta (\xi ) = (id \o \Delta )\Delta (\xi )\vp
       \ \ {\rm for\ all} \ \ \xi \in \G^* \ . \label{vpint}
\ee
\item
the following {\em pentagon  equation} holds
\be
   (id \o id \o \Delta )(\varphi )
   (\Delta  \o id \o id )(\varphi )
    = (e\o \varphi )      (id \o \Delta \o id )(\varphi )
    (\varphi \o e ) \ \ .   \label{phiphi}
\ee
\item  $\vp$ and the co-unit $\e$ satisfy
$(id \o \e \o id) (\vp)= \D (e)$.
\item there exist elements $\a, \b$ which satisfy relations
  (\ref{antip} f. ) together with the ``normalization''
\be \sum \S(\vp^1_\s)\a \vp^2_\s \b \S(\vp^3_\s) = e =
    \sum \phi^1_\s \a \S(\phi_\s^2) \b \phi^3_\s\ \ ,    \label{antonvp}
\ee
where $\phi^i_\s$ are defined through the expansion
$\vp^{-1} =  \sum \phi^1_\s \o \phi^2_\s \o \phi^3_\s$
\item $\vp$ is unitary, i.e. $\vp^* = \vp^{-1}$.
\end{enumerate}
An element $\vp \in \G^* \o \G^* \o \G^*$ with these properties
is called {\em re-associator}.
\end{defn}

Drinfel'd introduced the notion of quasi-co-associativity
in \cite{Dri2} for the case without
truncation, viz. $\D (e) = e \o e$.
Without truncation, $\vp^{-1}$ is a true inverse of $\vp$.
Let us discuss the meaning of this definition in terms of
representation theory. Consider tensor products of three
representations $\t^i,i=1,2,3$. Due to the freedom in placing the
brackets, there exist two different ways to perform threefold
tensor products,
$(\t^1 \bo \t^2)\bo \t^3 \ \ ${\em and }$ \ \
\t^1 \bo (\t^2\bo \t^3 ) $. They are
constructed from the two different combinations of the co-product,
$(\Delta \o id)\D$ resp. $(id \o \D)\D$.
If  $\D $ is quasi-co-associative in the sense
of the above definition, the two  threefold tensor products
are unitary equivalent. Indeed it follows from equations
(\ref{quasiinvvp},\ref{vpint}) and the
last item that $\vp$ furnishes an unitary
intertwiner  $(\t^1 \o \t^2 \o \t^3)(\vp)$. Definition
\ref{Defquasiass}.3 expresses equality of two intertwiners between
fourfold tensor products of representations. The name derives
from the fact that the equation describes commutativity of
a pentagon shaped diagram, in which the edges are indexed with
the five factors of the equation.
The relation in 4. is consistent since
$(\t^1 \bo \e) \bo \t^2 = \t^1 \bo\t^2 = \t^1 \bo (\e \bo \t^2)$
by triviality of $\e$.
Corresponding
equations for the other components of $\vp$ follow with the help of the
pentagon equation. In particular we will need the relation
\be
(id \o id \o \e ) (\vp) = \D (e)\ \ . \label{epsvp}
\ee

The definition above has another highly nontrivial consequence.
Given  a co-product $\D$ and an antipode $\S$ the combination
$(\S \o \S) \D'(\S^{-1}(\xi))$ defines again a homomorphism
$\G^* \mapsto \G^* \o \G^*$.
The latter turns out to be equivalent to the co-product
$\D$, if $\D$ is quasi-co-associative  \cite{Dri2}. To make this
statement more precise we introduce the following notation.
\bea
 \c &=& \sum \S(U_\s) \a V_\s \o \S(T_\s) \a W_\s \nn \\[1mm]
\mbox{with} & &                          \nn
\sum T_\s \o U_\s \o V_\s \o W_\s =
(\vp \o e)(\D \o id \o id)(\vp^{-1})\ \ , \\[2mm]
 f & = & \sum (\S \o \S)(\D'(\phi^1_\s))  \c   \D (\phi^2_\s \b
 \S(\phi^3_\s))\ \ , \label{fel} \\[1mm]
\mbox{with} & & \phi = \vp^{-1} = \sum \phi^1_\s \o
 \phi^2_\s \o \phi^3_\s\nn \ \ .
\eea
Drinfel'd proved in \cite{Dri2} that the element $f$ satisfies
\bea
 f \D(\xi) f^{-1}  & = &
 (\S \o \S) \D'(\S^{-1}(\xi)) \
 \ \mbox{ for all } \ \ \xi \in \G^*\ \ ,  \label{fint}  \\[1mm]
 \c & = & f \D (\a) \ \ . \nn
\eea
The first equation asserts that $f$ furnishes an intertwiner
between the coproduct $\D$ and the combination of $\D$ and
$\S$ on the right hand side. In terms of representation theory, it
can be restated as
\be
(\bt^I \bo \bt^J)(\xi) (\bt^I \o \bt^J)(f^*) =
(\bt^I \o \bt^J)(f^*) (\t^J \bo \t^I)^{-}(\xi) \ \ .
\label{frint}
\ee
Here $(\t^J \bo \t^I)^{-}$ is the conjugate of
$\t^J \bo \t^I$.

\begin{defn} {\em (quasi-triangularity) \cite{Dri2,MSIII}}
 \label{Defquasitri}
 A  bi-*-algebra with antipode
 $(\G^*,\D,\e,*,\S)$ with quasi-co-associative co-product $\D$
 and re-associator $\vp$
 is called quasi-triangular,
 if there exists $R \in \G^* \o \G^*$ such that
\begin{enumerate}
\item
 $R$ has a quasi--inverse
 $R^{-1} \in \G^* \o \G^*$
 such that
 \be
  R R^{-1} = \D ' (e)\ \ \ \ \ \ ,
  \ \ \ \ \ \  R^{-1} R = \D(e)\ \ .
 \ee
\item
 $R$ satisfies the ``intertwining relation''
 \be   R \D(\xi) = \D ' (\xi) R \ \ \mbox{ for all } \ \ \xi \in \G^*
 \ \ .\label{Rint} \ee
\item $R$ is unitary in the sense that $R^*= R^{-1}$.
\item
the following {\em hexagon equations} are fulfilled
\bea
   (id \o \Delta )(R) &=& \varphi_{231}^{-1} R_{13}\varphi_{213}R_{12}
  \varphi^{-1} \ ,    \nn
  \\
   (\Delta \o id  )(R) &=& \varphi_{312} R_{13}\varphi_{132}^{-1}R_{23}
  \varphi \ .            \label{hex}
\eea
\end{enumerate}
We used the standard notation. If $R = \sum r_a^1 \o r_a^2$  then
$R_{13} = \sum r_a^1 \o e \o r^2_a $ etc. Given the expansion
$\varphi = \sum \varphi_{\sigma}^1
\o \varphi_{\sigma}^2 \o \varphi_{\sigma}^3$ and any permutation
s of 123 we set
$\varphi_{s(1)s(2)s(3)} = \sum_{\sigma} \varphi_{\sigma}^
{s^{-1}(1)} \o \varphi_{\sigma}^{s^{-1}(2)} \o \varphi_{\sigma}
^{s^{-1}(3)} $ .
\end{defn}

\noindent
The discussion of this definition parallels the one given for
definition \ref{Defquasiass}.
Quasi-triangularity implies that the
two representations $\t^1 \bo \t^2$ and $\t^2 \bo \t^1$ are
equivalent. The intertwiner is furnished by
$(\t^1 \bo \t^2)(R)$. Equations $(\e \o id)R = e, (id \o \e)R=e
$ follow with the help of the hexagon equation. The same holds
true for all relations involving the action of the antipode $\S$
on components of $R$.

After these definitions we are prepared to explain the title
of this section. A bi--*--algebra with antipode $\S$,
re-associator $\vp$ and $R$-element $R$ is called
weak quasitriangular quasi-Hopf-*-algebra or simply
{\em weak quasi quantum group} \cite{MSIII}.
They are generalizations of Drinfeld's quasi quantum
groups \cite{Dri2} in which
truncation is not allowed, i.e. $\D (e) = e \o e$.
For a quantum group \cite{Dri1,Wor1,Wor2},
 the re-associator $\vp$ is trivial,
i.e. $\vp =e \o e \o e$. In this framework group
algebras appear as special examples of quantum groups
when $R = e \o e$.

Let us discuss some properties of
bi-*-algebras with antipode $\S$, re-associator $\vp$
and $R$-element $R$.
The relations stated in definition \ref{Defquasiass},\ref{Defquasitri}
imply validity of quasi
Yang Baxter equations,
  \be
  R_{12}\varphi_{312}R_{13}\varphi^{-1}_{132}R_{23}\varphi
  = \varphi_{321}R_{23}\varphi^{-1}_{231} R_{13}\varphi_{213}R_{12}
  \ , \label{QYBE}
  \ee
and this guarantees that $R$ together with $\varphi$ determines
a representation of the braid group \cite{MSVI}.
To state this result we introduce some notations. Write
\be
e^n = e\o \dots \o e \ \ \ \ \ \mbox{(n factors)} \label{notations}
\ee
and similarly for $id^n$. In addition we abbreviate
$   {\G^*}^{\o n}= \G^* \o ... \o \G^*$          (n factors),
and
\bea
\Delta^n &=& (id^{n-1} \o \Delta ) \cdots (id \o \Delta ) \Delta
 \ \ \mbox{ for } \ n \geq 2 \ , \\
  \Delta^1 &=& \Delta \ , \ \Delta^0 = id \ , \ \Delta^{-1}= \e \ .
\eea
Furthermore we introduce the following permutation maps
${\bf P}^n_k : \G^{*\o n} \mapsto \G^{*\o n}$ defined by
\be
{\bf P}^n_k (\xi_n \o \dots \xi_{k+1} \o \xi_{k} \dots \o \xi_1)
 = ( \xi_n \o \dots \xi_{k} \o \xi_{k+1} \dots \o \xi_1) \ \ .
\ee

\begin{theo} {\em (Artin relations) \cite{MSV,MSVI}}  \label{braidinqqg}
Let $R^+= R$ and
$ R^- =  {R'}^{-1}$ where $'$ interchanges factors in
$\G^* \o \G^*$. For $k = 1, \dots ,n-1$
define maps $\s^{n\pm}_k: \G^{*\o n} \mapsto \G^{*\o n}$ by
\be \sigma_k^{n\pm}=  \D ^{n-1}(e) {\bf P}^n_k
 (id^{n-k+1} \o \D^{k-2})(e^{n-k-1}
\o \vp_{213} ( R^{\pm}\o e)\vp^{-1})\ .
\ee
Then $\s^{n\pm}_k$
obey Artin relations (\ref{Artinu}) and
$\s^{n-}_k$ is the quasi-inverse of $\s^{n+}_k$, i.e.
$$  \s^{n+}_k \s^{n-}_k = \D^{n-1}(e)\ \ .$$
\end{theo}

The proof can be found in \cite{MSVI}. For the special case
in which $\vp= e \o e \o e$ and $\D(e) = e \o e$ the
explicit relation to representations of the braid group
prompted the discovery of
quantum groups and can be used to construct
them from solutions of the Yang Baxter equations (see
e.g. \cite{Maj1}).

Let ${\tau^I}$ denote a complete set of
representatives from the equivalence classes of irreducible
representations of $\G^*$.
$\t^I$ are assumed to be finite dimensional and
unitary representations on the Hilbert spaces $V^I$, i.e.
$\tau^I(\xi^*) = (\tau^I(\xi))^*$.
For semisimple $\G^*$,
representations $\t^I \bo \t^J$ can be decomposed into
a direct sum of representations $\t^K$ and this decomposition
determines a Hilbert space $\C(IJ,K)$ of Clebsch Gordon
intertwiners
$C(IJ \Y K):V^I \o V^J \to V^K$ as in section 3.3.

When a re-associator $\vp$ and an element $R$ exist, the tensor product
of representations is associative and commutative up to equivalence so
that the dimensions $\nu^{IJ}_K= dim \C (IJ \Y K)$ furnish a
solution of the eqs. (\ref{comass}).
To describe the action of $\vp$ and $R$ on the
Clebsch Gordon intertwiners, we fix a set of
complex phases $\omega^{IJ} = \overline {\omega^{JI}}$
such that
\be
\omega^{IJ} \omega^{IK} = \omega^{IL} \ \ \ , \ \ \
\omega^{JI} \omega^{KI} = \omega^{LI}  \label{omegaprop}
\ee
whenever $\t^L$ is a subrepresentation of $\tau^J \bo \tau^K$.
Note that one can always choose $\omega^{IJ} = 1$ for all $I,J$.
Due to the intertwining properties (\ref{vpint},\ref{Rint})
of $\vp$, $R$  and Schur's lemma, $\vp , R$ determine
a set of complex matrices $\Phi,\omega$ defined by
\bea
C(IP \Y L) C(JK \Y P)_{23} \vp^{IJK} &=& \sum_Q \FS{P}{Q}{I}{J}{L}{K}^*
C(QK \Y L) C(IJ \Y Q)_{12} \ \ ,\label{pathvp}\\
C(IJ \Y K) \hat R^{+IJ}  &=& \omega^{IJ} \OS{K}{I}{J} C(JI \Y K)\
\label{pathR} \ \ .\eea
Here $\vp^{IJK} = (\t^I \o \t^J \o \t^K) (\vp)$,
$\hat R^{+IJ}= P (\t^I \o \t^J)(R)$ with
$P: V^I \o V^J \mapsto V^J \o V^I $
the permutation  map.
The properties of $R, \vp$ give rise to relations
among the matrices $\Phi,\omega$. A short calculation shows
that they satisfy the polynomial equations
(in proposition \ref{poly}) when $\Phi$ substitutes for $F$
and the matrix $\omega$ appears in place of $\Omega$.

Conversely we can start from a bi-*-algebra $\G^*$ with antipode
with an associative and commutative set of dimensions
$\nu^{IJ}_K \equiv ${\it dim\/}$\C(IJ \Y K)$. Note that the
dimensions of the matrices $F,\Omega $ are the only
parameters in the polynomial equations. In proposition
\ref{poly} these dimensions were given by the fusion
rules $N^{IJ}_K$. However, it is realized immediately
that every solution of (\ref{comass}) -- in particular
$\nu^{IJ}_K$ --  determines a set of polynomial equations.
Our aim is to construct elements $\vp$ and $R$
from a known solution of these polynomial equations.
We will succeed for semisimple algebras $\G^*$.

\begin{theo} {\em (reconstruction theorem)} \label{wqqgconstr}
Let $\G^*$ be a semisimple bi-*-algebra with antipode.
Suppose that the multiplicities $\nu^{IJ}_K$ which appear
in the Clebsch Gordon decomposition $$\t^I \bo \t^J \cong
\bigoplus \nu^{IJ}_K \t^K$$ are commutative and associative
in the sense of eq. (\ref{comass}) and that a solution
of eq. (\ref{omegaprop}) has been fixed. Then
every solution of the polynomial equations
(proposition {\em \ref{poly}}) associated with
$\nu^{IJ}_K$ determines a pair of elements $\vp,R $
with properties as in definition {\rm \ref{Defquasiass},
\ref{Defquasitri}}. Their action on Clebsch Gordon maps
is given by
eq. (\ref{pathvp},\ref{pathR}).
\end{theo}

{\sc Proof:} The proof of this theorem consists of two parts. First
one has to show that for a given set of $\vp^{IJK},
  \hat R^{+IJ}$ defined
by  (\ref{pathvp},\ref{pathR}) one can always find elements
$\vp \in \G^* \o \G^* \o \G^*$ and $R \in \G^*\o \G^*$ such that
$\vp^{IJK} = (\t^I \o \t^J \o \t^K) \vp $, $\hat R^{+
 IJ} = P(\t^I \o \t^J ) ( R)$.
We do this for $R$. Let $M_K$ be the full matrix algebra that
consists of $(dim(\t^K) \ti dim(\t^K))$-matrices. Since $\G^*$
is semisimple, $\t^K(\G^*) = M_K$ for all irreducible
representations $\t^K$. By definition
$P\hat R^{+ IJ} \in M_J \o M_I$ so that $P\hat R^{+ IJ}$ is a sum of
tensor products of matrices, $P\hat R^{+IJ} = \sum_{\s} m_{\s}^1
\o m_{\s}^2$, $m^1_{\s} \in M_J, m^2_{\s} \in M_I$.
We can find elements $s^i_{\s}\in \G^*, i=1,2,$ such that
$\t^J(s^1_{\s})=m^1_{\s}$ and $\t^I(s^2_{\s})=m^2_{\s}$.
Take $S^{JI} \in \G^*$ to be the element $P^J \o P^I \sum
s^1_{\s} \o s^2_{\s}$ ($P^I$ is the minimal central projection
corresponding to the irreducible representation $\t^I$)
 and repeat this construction for every
pair $(I,J)$ of representations. Finally, $R = \sum_{IJ} S^{
IJ}$ satisfies $\hat R^{+IJ} = P(\t^I \o \t^J)( R)$.
The second part of the proof is to show that the elements
$\vp$,$R$ satisfy all the relations in Definition
(\ref{Defquasiass},\ref{Defquasitri}). This is an immediate
consequence of the polynomial equations
(proposition \ref{poly}). Proving the normalizations
(\ref{antonvp}) involves an appropriate choice of $\a,\b$
by exploiting the transformations $\a \mapsto \zeta \a$
described after relations (\ref{staranti}).
The quasi-inverses act according to
\bea
C(QK \Y L) C(IJ \Y Q)_{12} (\vp^{-1})^{IJK} &=& \sum_P \FS{P}{Q}{I}{J}{L}{K}
C(IP \Y L) C(JK \Y P)_{23} \ \ ,\\
C(JI \Y K) \hat R^{-IJ}  &=& \omega^{JI} \OS{K}{I}{J}^* C(IJ \Y K)\ \ ,
\eea
with $(\vp^{-1})^{IJK} =( \t^I \o \t^J \o \t^K)(\vp^{-1})$,
 $\hat R^{-IJ} = (\tau^J \o \tau^I ) (R^{-1}) P$ and $P:
V^I \o V^J \mapsto V^J \o V^I$ the permutation map.

The idea to reconstruct elements $\vp$ and $R$ from a
solution of the polynomial equations appeared first in \cite{MSIV}.
There is was done in a concrete example. In the language
of categories, a similar observation was formulated by
Majid \cite{Maj2} and extended to cases with truncation
by Kerler in \cite{Ker1}.

\subsection{Construction of weak quasi quantum groups}

Looking for examples of weak quasi quantum groups, the
last theorem in the preceding subsection gives a simple
strategy. Since solutions of the polynomial equations
are known for many fusion rules $\nu^{IJ}_K$, we obtain a
weak quasi quantum group whenever we are able to find
a suitable bi-*-algebra with antipode (in the sense of theorem
\ref{wqqgconstr}). The construction of bi-*-algebras with antipode
with prescribed multiplicities $\nu^{IJ}_K$ of the Clebsch
Gordon decomposition is actually a simple algebraic
problem. It can be solved if $\d_I \d_J \geq \sum \nu^{IJ}_K
\d_K$ has a positive integer solution.

As usual, the multiplicities $\nu^{IJ}_K$ are assumed to
be associative and commutative in the sense of
eq. (\ref{comass}). We suppose that there a unique label $0$
with the property $\nu^{I0}_J = \delta_{I,J}$ and that
-- with respect to  $0$ -- every label $J$ has a
conjugate $\bar J$. The latter is distinguished
by the property $nu^{IJ}_0= \delta_{I \bar J}$.

In order to start our construction we choose a
set of finite dimensions $\delta_J \geq 1$ such that
$\delta_0 = 1$ and $\delta_J = \delta_{\bar J}$.
$V^J$ should denote a $\delta_J$--dimensional
Hilbert space and $e_J$ the unit operator on $V^J$.
$V^0$ is spanned by one normalized vector $\inv$ with
dual $\vni$.

For certain choices of the dimensions
$\delta_J$ we can hope to find a family of linear maps
$C^a(IJ \Y K): V^I \o V^J \mapsto V^K
( a =  1 \dots N^{IJ}_K)$ which
satisfies the following equations.
\bea
\label{orthonorm}
C^a(IJ \Y K) C^b(IJ \Y L)^*  & = &  \delta_{K,L} \delta_{a,b} e_K \ \ ,\\[1mm]
\label{norma}
C(I0 \Y I) = & e_I & = C(0I \Y I)\ \ , \\ [1mm]    \label{anti}
C(\bar I I  \Y  0)_{23} C(I \bar I  \Y 0)_{12}^* & = & \kappa_I^{-1} e_I\ \ .
\eea
Here, the constants $\kappa_I$ are supposed to be real numbers.
We introduce the Hilbert space $V= \bigoplus_J V^J$ and
extend $e_J$ and $C^a(IJ \Y K)$ to linear maps on $V$ resp.
$V\o V$ by $e_I V^J = e_I \delta_{I,J}$ etc..

\begin{prop} \label{Hopfalg}
Suppose that equations (\ref{orthonorm},
\ref{norma}) can be solved
by a family $C^a(IJ \Y K)$. Define a *-algebra $\G^*$
which consists of maps $\xi: V \mapsto V$ as follows
$$ \G^* = \{ \xi :V \mapsto V  \Y  \xi : V^I \mapsto V^I
   \mbox{ for all } I \in \I \}\ . $$
The *-operation on $\G^*$ is given by the usual adjoint
of maps $\xi: V \mapsto V$.
There exist a co-product $\D$, a co-unit $\e$ and
an antipode $\S$ which enjoy the usual properties.
Explicitly they act on elements $\xi \in \G^*$ as
\bea
\D(\xi) &=& \sum C^a(IJ \Y K)^*  \xi C^a(IJ \Y K)\ \ , \\
\e(\xi) &=& \vni \xi \inv \ \ ,\\
\S(\xi) &=& \sum_{I} \kappa_I
C(\bar I I \Y 0)_{23}  (e_I \o \xi \o e_I) C(I \bar I  \Y 0)_{12}^*\ \ .
\eea
\end{prop}

{\sc Proof:} Most of the properties are obvious. We shall
give only some of the calculations and omit the others.
Orthonormality (\ref{orthonorm}) of $C^a(IJ \Y K)$ is needed for
the co-product $\D:\G^* \mapsto \G^* \o \G^* $
to become a homomorphism.
\bea
\D (\xi) \D(\eta ) & = & \sum C^a(IJ \Y K)^* \xi C^a(IJ \Y K)
    \sum C^b(LM \Y N)^* \eta C^b(LM \Y N) \nn \\
                   & = & \sum C^a(IJ \Y K)^* \xi \delta_{a,b}
    \delta_{K,N} e_K \eta C^b(IJ \Y N) = \D(\xi \eta )\ \  . \nn
\eea
$\D (\xi )^* = \D (\xi^*)$ and the properties of $\e$
are trivial. The normalization (\ref{norma}) is used to obtain
\be (\e \o id ) \D (\xi) = \sum C(0J \Y J)^* \xi C(0J \Y J) = \xi
\ . \nn \ee
To proof that $\S:\G^* \mapsto \G^*$ is an
anti-homomorphism we note that the definition of $\S$ is equivalent
to
\be
C(I\bar I \Y 0)(e \o \S (\xi )) = C(I \bar I \Y 0) (\xi \o e )
\label{help}\ \ .  \ee
Consequently, the action of $\S$ on the product $\xi \eta$ can be
evaluated according to
\bea
\S(\xi \eta) & = & \sum_{I} \kappa_I C(\bar I I \Y 0)_{23}
                   (e_I \o \xi \eta \o e_I)
                   C(I \bar I  \Y 0)_{12}^* \nn \\
             & = & \sum_{I} \kappa_I C(\bar I I \Y 0)_{23}
                   (e_I \o  \eta \o \S(\xi))
                   C(I \bar I  \Y 0)_{12}^* \nn \\
             & = & \S(\eta) \S(\xi)
\nn \eea
The behaviour of $\S$ with respect to the *-operation is checked
as follows.
\bea
\S(\xi )^* & = & \sum_{I} \kappa_I C(\bar I I \Y 0)_{12}  (e_I \o  \eta^* \o
e_I
)
                   C(I \bar I  \Y 0)_{23}^* \nn \\
           & = & \sum_{I} \kappa_I C(\bar I I \Y 0)_{23}
                 (\S^{-1}(\xi^*) \o  e_{\bar I} \o e_I)
                 C(I \bar I  \Y 0)_{23}^* \nn \\
           & = & \S^{-1} (\xi^*) \sum_I \kappa_I \kappa_I^{-1}
                 = \S^{-1}(\xi^*) \nn\ \ .
\eea
It remains to prove compatibility with the co-product. To do this
we start again from equation (\ref{help}).
\bea
\S(\xi^1_\s) \xi^2_\s & = & \sum_I \kappa_I C(\bar I I  \Y 0)_{23}
                      C(I \bar I \Y 0)^*_{12} \S(xi^2_\s)\xi^1_\s \nn \\
                      & = & \sum_I \kappa_I C(\bar I I \Y 0)_{23}
                      (e \o \D(\xi)) C(I \bar I  \Y 0)^*_{12} \nn\\
                      & = & \sum_I \kappa_I \e(\xi) C(\bar I I \Y 0)_{23}
                      C(I \bar I  \Y 0)^*_{12} \nn \\
                      & = & \e(\xi)\ \ .
\eea
A similar calculation gives $\xi^1_\s \S(\xi^2_\s) = \e(\xi)$. We see
that in these examples $\a=\b=e$. Triviality of $\a,\b$ is related to
the condition (\ref{anti}). If $\k_I$ is allowed to become a
more general invertible map $V^I \mapsto V^I$, one has to encounter
nontrivial $\a,\b$.

Irreducible unitary representations $\t^J$ of
$\G^*$ are obtained by restriction to $V^J$.
$$ \t^J(\xi) \equiv \xi \Y _{V^J}\ \ . $$
The tensor product $\t^I \bo \t^J$ acts
on $V^I \o V^J $ by
$$
(\t^I \bo \t^J)(\xi) =  \sum_{K,a}
C^a(IJ \Y K)^* \t^K(\xi) C^a(IJ \Y K)\ \
$$
so that $C^a(IJ \Y K)$ have been identified as
Clebsch Gordon intertwiners. By construction,
{\it dim}$(\C(IJ \Y K))= \nu^{IJ}_K$. We conclude
that our initial problem is solved by
the bi-*-algebra with antipode $(\G^*,\D,\e,*,\S)$ in
proposition \ref{Hopfalg},
provided it exits.

The above proposition reduces the problem
of finding an appropriate bi-*-algebra with antipode to
the solution of equations (\ref{orthonorm} f.). Of course
such solutions will only exist for special
choices of the dimensions $\delta_J$. A
necessary condition on $\delta_J$ can easily
be derived from the orthonormality of
$C^a(IJ \Y K)$ (\ref{orthonorm}). It means that the $C^a(IJ \Y K)$
map vectors in the $\delta_I \delta_J$--dimensional
Hilbert space $V^I \o V^J$ onto an orthogonal sum
of $\delta_K$--dimensional spaces which occur with a
multiplicity $N^{IJ}_K$ depending on K. The total
dimension $\sum _K N^{IJ}_K \delta_K$ of the
image cannot exceed the dimension $\delta_I \delta_J$
of $V^I \o V^J$,
\be \delta_I \delta_J \geq \sum_K N^{IJ}_K \delta_K\ \ . \label{neccon}
\ee
The first part in the proof of the following corollary
shows that condition (\ref{neccon}) is also sufficient.

\begin{coro} \label{Ex}
Suppose that the number of labels $I$ is finite
and that the ``fusion rules'' $\nu^{IJ}_K$ satisfy the
standard assumptions. Then there is a semisimple bi-*-algebra
$\G^*$ with antipode and
irreducible unitary representations $\t^I$ such that
\be \mbox{ dim\/}(\C(\t^I \t^J \Y  \t^K)) = \nu^{IJ}_K \ \ . \ee
\end{coro}

{\sc Proof: } Let us first suppose that a solution of
(\ref{neccon}) exists. One starts by constructing
maps $C(I \bar I \Y 0 ): V^I \o V^{\bar I} \mapsto V^0 $ with
the property that $C(I \bar I  \Y 0)_{23} C(\bar I I  \Y 0)_{12}^*$
is proportional to $e_I$.
They will satisfy relations (\ref{orthonorm}, \ref{anti})
after an appropriate normalization.
A complete set of solutions
of (\ref{orthonorm},\ref{norma},\ref{anti}) can then be
obtained by the usual orthonormalization procedure.
Relation (\ref{neccon}) guarantees that  sufficiently
many linear independent maps exist.

For finite number of equivalence classes of irreducible
representations, a solution of (\ref{neccon}) is furnished by
$\d_0 = 1$ and $\d_K = \delta= \mbox{\it max\/}_{IJ}( \sum_K N^{IJ}_K)
, J \neq 0,$. This proves the corollary.

In the context of this subsection, corollary \ref{Ex} asserts the
existence of (non-trivial) weak quasi quantum groups. It should be
remarked that it also answers the question which was raised by the
assumption (\ref{fundAss}) at the beginning of section 3.3., namely
the question whether a bi-*-algebra $\G^*$ with antipode suitable to
construct covariant field operators $\Psi^J_j(\rho_J)$ does exist.
As is seen from corollary \ref{Ex}, one can find appropriate
bi-*-algebras with antipode  for all rational models.

\noindent
{\bf Example:} Let us describe at least one explicit solution
of the equations (\ref{orthonorm} f.) for the fusion rules
$N^{IJ}_K$ of the chiral critical Ising model.
They can be found in  section 7. Since the dimensions
$\delta_0 =1, \delta_{1/2}=2, \delta_1 =1$ satisfy the
condition (\ref{neccon}), we know that a bi-*-algebra with antipode
exists for these dimensions. A list of corresponding Clebsch
Gordon maps which satisfy eqs. (\ref{orthonorm},\ref{norma},\ref{anti})
is given by (with {\bf 1} the two--dimensional unit matrix)
$$ \begin{array}{rll}
C(00 \Y 0)&=1   & C(\frc \frc  \Y  0) = {\scriptstyle \frac{1}{\sqrt{2}}}
              (1\ 0\ 0\ 1) \\
C(0\frc \Y  \frc )& = {\bf 1}
         & C(11 \Y 0) = 1 \\
C(01 \Y 1)&=1  & C(\frc \frc  \Y  1) = {\scriptstyle \frac{1}{\sqrt{2}}}
              (1\ 0\ 0\ -1) \\
C(\frc 0 \Y \frc)& = {\bf 1}
          &  C(10 \Y 1)=1  \\
C(\frc 1 \Y \frc)& = \mbox{\it diag\/}(1,-1)\ \ \ \ \ \ \ \ \ \
          &
C(1 \frc  \Y \frc) = \mbox{\it diag\/}(1,-1)
\end{array} $$
Here $(1\ 0\ 0\ 0)=(1\ 0 ) \o (1\ 0),\ (0\ 1\ 0\ 0)=(1\ 0)\o (0 \ 1)$
etc.. A solution of the polynomial equations associated with
the Ising fusion rules $N^{IJ}_K$ is well known and determines
a re-associator $\vp$ and a $R$-matrix $R$ according to
equations  (\ref{pathvp},\ref{pathR}). So we obtain the (smallest
non-trivial) example of a weak quasi quantum group.

\section{Quantum Symmetry, Statistics and Locality}
\setcounter{equation}{0}

In
quantum field theory, permutation group statistics is
implemented through
quadratic relations among the field operators, viz. canonical
(anti--)com\-mu\-tation relations for Bosons (Fermions).
The spin statistics theorem states that Fermions have spin
$s = \frac{1}{2}, \frac{3}{2}, \dots $,
whereas Bosons have integer spin. More general values for the spin
(remember that the spin
 labels representations of the rotation group, e.g. SO(2) in
2 space dimensions)
are possible in low dimensional quantum field theory. They are
associated with braid group statistics.
It has been proposed to implement braid group statistics through
local braid relations \cite{Fro2}.
\be        \label{LBR}
\Psi^I_i (x,t) \Psi^J_j (y,t)  = \omega^{IJ}
 \Psi^J_l (y,t) \Psi^I_k (x,t)
\R^{IJ>}_{kl,ij}  \ \ \ \mbox{ for } \ \    x>y \ . \ee
Here, $\omega^{IJ}$ are complex phase factors.
As in section 2, the meaning of $x>y$ depends on the dimension
of the space-time on which the theory lives.
In contrast with \cite{Fro2} we
do not restrict the
$\R$-matrix to have {\bf C}--number entries, but the matrix
elements may take values in $\U(\G^*)$ instead.
For the order $x<y$ a similar relation follows with
$ \R^{JI<} $ and $\omega^{JI} = \overline{\omega^{IJ}}$.
Denoting with $\hat \R$ the matrix obtained
from $\R$ by interchange of the first indices $k,l$, the
two matrices $\R^{IJ>}$ and $\R^{JI<}$ obey
$$ \hat \R^{IJ<} = (\hat \R^{JI>})^{-1}\ \ .$$
Note that for $\R^{IJ>}_{kl,ij} = 1 $ and $\omega^{IJ}
= \pm $ we recover
Bose/Fermi--commutation relations as
a special case of eq. (\ref{LBR}).

Consistency of local braid relations with the transformation
law (\ref{covtrans}) distinguishes weak quasi quantum
groups from arbitrary bi-*-algebra symmetries
\cite{MSIII}. We will review these
arguments here. In the second subsection
we determine a re-associator $\vp$ and a
$R$-element $R$ for the quantum symmetry of
the algebra of fields $\B$ constructed in
section 3.3. Local braid relations of the field
operators (\ref{Fieldop}) and their covariant adjoints
will be established with
$\R^{IJ>}_{kl,ij}$  furnished
by the elements $\vp,R$.

\subsection{Weak quasi quantum group symmetry and local braid
            relations}

In a quantum theory with quantum symmetry
the matrix-elements $\R^{IJ>}_{kl,ij}$ are
not free to take any values. Consistency of
(\ref{LBR}) with the transformation law of field operators and
associativity of the product of operators constrain $\R$.
If the quantum symmetry is
quasi-co-associative and quasi-triangular with re-associator $\vp$
and $R$-element $R$, a solution of these constraints is given by
\bea        \nn
 \R^{IJ>}_{kl,ij}  & = &  (\tau^I \o \tau^J \o \U )_{kl,ij}
 (\varphi_{213} (R \o e) \varphi^{-1})
\label{Rmatrix} \\
 & = &  \sum
 \tau^I_{ki}(\vp^2_\s r_a^1 \phi_\t^1)
 \tau^J_{lj}(\vp^1_\s r_a^2 \phi_\t^2)
 \U(\vp^3_\s \phi_\t^3 ) \ ,
\label{RSQ}
\eea
where we use the same notations as in the preceding
section and $\phi = \vp^{-1}$. From the second line we read off that
the matrix elements $\R^{IJ>}_{kl,ij}$ are not numbers in general
but operators in the Hilbert space. The expression
is a linear combination of representation operators
$ \U(\vp^3_\s \phi_\t^3) $
and is therefore equal to a representation operator $U(\eta)$ of
some element $\eta \in \G^* $. A numerical matrix $\R$ is
obtained if and only if the re-associator $\vp$ is trivial, i.e.
if $\vp = e \o e \o e$.

To motivate eq. (\ref{Rmatrix})
we demonstrate that {\it local braid relations
{\rm (\ref{LBR})} with $\R^{IJ>}$ given by {\rm (\ref{Rmatrix})}
are  consistent
with the transformation law of fields}, i.e. that both sides of the
equation transform in the same way. The
products  of
covariant fields which appear in (\ref{LBR}) are in general
not covariant, if the co-product $\D$ is not
co-associative. Suppose now that there exists a
re-associator $\vp$ satisfying (\ref{vpint}). Then one can
construct a
``covariant product'' $\ti$ of field operators
\cite{MSIII,MSV}.

\begin{defn}{\em (covariant product of field operators)}\label{COVPF}
The {\em covariant product} of two multiplets $\Psi^I,\Psi^J$
is the multiplet defined by
\be
 (\Psi^I \ti \Psi^J )_{ij} \equiv  \sum
 \Psi^I_m \Psi^J_n \t^I_{mi}(\vp_\s^1)
 \t^J_{nj}(\vp_\s^2)  \U(\vp^3_\s)  \ \ .
\label{CovProd}
\ee
\end{defn}

By  (\ref{vpint}), {\em $
\Psi^I \ti \Psi^J$ transforms covariantly according
to the tensor product representation  $\t^I \bo \t^J$}.
Ordinary products of field operators can be recovered
from covariant ones,
\be \Psi^I_i\Psi^J_j = \sum (\Psi^I\ti  \Psi^J )_{mn}
 \t^I_{mi}(\phi^1_{\s })
 \t^J_{nj}(\phi^2_{\s})
  \U(\phi^3_{\s })   \ . \label{Finv} \ee
Using this covariant product,
local braid relations
(\ref{LBR},\ref{Rmatrix}) may be rewritten as
\be
 (\Psi^I \ti \Psi^J )_{ij} = \omega^{IJ} (\Psi^J\ti \Psi^I)_{lk}
 (\t^I_{ki} \o \t^J_{lj})(R) \ \ .  \label{Rtilde}
\ee
The arguments of the field operators, which we neglect to write,
satisfy conditions as in eq.(\ref{LBR}).
Because of the
covariance properties of field operators, both sides transform
in the same way if
the element $R\in \G^*\o \G^*$
obeys intertwining property (\ref{Rint}).

To prepare for our second consistency check one should
notice that associativity of the product of field
operators is equivalent to the following
\begin{prop} \label{QASSF}
              {\em (quasi-
associativity of the covariant product)} The covariant product
is quasi-associative in the sense that products $\Psi^{J_n}_{j_n}
\ti \dots \ti \Psi^{J_1}_{j_1}$ with arbitrary positions
of brackets can be written as a complex linear combination of
products $\Psi^{J_n}_{k_n} \ti \dots \ti \Psi^{J_1}_{k_n}$
with any other specification of brackets. In particular one has
\be
 ((\Psi^I\ti \Psi^J)\ti \Psi^K)_{ijk} =
 (\Psi^I\ti (\Psi^J\ti \Psi^K))_{i'j'k'}
 (\t^I\o \t^J\o \t^K)_{i'j'k',ijk}(\vp )\ .
\label{QAss}
\ee
\end{prop}

{\sc Proof:}This proposition is a consequence of
the pentagon equation (\ref{phiphi}) for $\vp$.

Consistency requires that the matrix $\R^{IJ>}$ satisfies
constraints which come from the possibility of interchanging triples
$\Psi^I_i(x,t) \Psi^J_j(y,t) \Psi^K_k(z,t)$, $x>y>z$,
of fields in two different ways, leading to the same result
where fields in the multiplet $\Psi^K$ appear in the leftmost
position followed by $\Psi^J$ and finally $\Psi^I$ to the
right.
The constraints are exploited most easily if one starts
from the threefold $\ti$--product $\Psi^I \ti (\Psi^J \ti \Psi^K)$.
By an alternating application of eq. (\ref{QAss}) and
eq. (\ref{Rtilde}) we end up with $(\Psi^K \ti \Psi^J) \ti \Psi^I$.
There are two ways to perform these manipulations which
give the same result, provided that
$R$, $\vp$ satisfy the quasi Yang Baxter equations (\ref{QYBE}).

In our present context
eqs. (\ref{hex}) imply validity of local braid
relations for composite operators, i.e. when the product
$\Psi^{I_1} \times \Psi^{I_2}$ is inserted in place
of $\Psi^I$. More precisely suppose that local
braid relations (\ref{Rtilde})
hold for the products $\Psi^{I_i}(x_i,t) \ti
\Psi^{J}(y,t), i=1,2$ and $x_i > y$. Then by (\ref{hex}) we have
the following {\em local braid relations for composites}
$$
((\Psi^{I_1} \ti \Psi^{I_2}) \ti \Psi^J)_{i_1i_2j} = \omega^{I_1J}
\omega^{I_2J}
(\Psi^J \ti (\Psi^{I_1} \ti \Psi^{I_2}))_{lk_1k_2}
((\tau^{I_1} \bo \tau^{I_2})_{k_1k_2i_1i_2}\o \t^J_{lj})(R)
 \ \ .$$
Similar considerations for the field operator $\Psi^J$
lead to
$$
(\Psi^{I} \ti (\Psi^{J_1} \ti \Psi^{J_2}))_{ij_1j_2} = \omega^{IJ_1}
\omega^{IJ_2}
((\Psi^{J_1} \ti \Psi^{J_2}) \ti \Psi^{I})_{l_1l_2k}
(\t^I_{ki} \o (\tau^{J_1} \bo \tau^{J_2})_{l_1l_2j_1j_2} )(R)
\ \ .
 $$
For the proof one uses the quasi--associativity (\ref{QAss})
of the $\ti$ product and local braid relations (\ref{Rtilde})
as in the preceding paragraph.

When we defined the covariant conjugation  \footnote{It is
now regarded as a complex anti-linear
map which sends every covariant multiplet
of fields to its covariant adjoint} in section 2, we
had to make the assumption (\ref{vpiint}) about the existence of
an appropriate intertwiner $\vp$. If $\vp$ is the re-associator
of a weak quasi-quantum group, the covariant conjugation has
certain distinguished properties. We state them here without
proofs. For details the reader should consult appendix A.
Recall that the covariant adjoint of a multiplet $\Psi^I_i$
transforming covariantly according to some representation
$\t^I$ of $\G^*$ was defined by ( rel. (\ref{covad}) )
\be
\br\Psi^I_i = (\Psi^I_j (\t^I_{ji} \o id)(w) )^* .
\ee
$\br\Psi^I_i$ transforms covariantly according to the
representation $\bt^I$ so that its covariant adjoint is
again well defined.
\be
 \br{\br\Psi^I_i} = (\br\Psi^I_j (\bt^I_{ji} \o id) (w) )^*
 =  \Psi^I_i
\ee
The last equality follows with the help of the pentagon equation
(\ref{phiphi})  for $\vp$. It states that the covariant conjugation
is involutive. Being kind of a substitute for the ordinary
adjoint $*$ (for $\vp = e \o e \o e$ the covariant adjoint is
given by $\br\Psi^I = \Psi^{I*}$), we expect the covariant
conjugation to be consistent with the product of fields. The
precise statement is
\be
 \overline{ (\Psi^I \ti \Psi^J)}_{ij} =
 (\br\Psi^J \ti \br\Psi^I)_{kl}
 (\bt^J_{kj} \o \bt^I_{li})(f^*) \ \ .
 \label{conjprod}
\ee
The term involving $f^*$ should not surprise. In view of relation
(\ref{frint}) it can be understood from consistency with the
transformation properties of relation (\ref{conjprod}).
As a last property of covariant conjugation we want to mention that
the adjoints $\br\Psi^I, \br\Psi^J$ of covariant fields
obey local braid relations
\be \label{conjbraid}
\br\Psi^I_i\  \br\Psi^J_j  =  \omega^{IJ}\
\br\Psi^J_l \ \br\Psi^I_\k \ (\bt^I_{ki} \o \bt^J_{lj}\o \U) (\R)\ \ ,
\ee
if $\Psi^I,\Psi^J$ satisfy (\ref{LBR},\ref{Rmatrix}).

The local braid relations for composites together with
relations (\ref{conjbraid}, \ref{conjprod})
reveal a remarkable {\em stability of local braid relations
with respect to covariant multiplication and conjugation}.
This has a major conceptual significance. Suppose we were able
to construct a set of ``fundamental'' covariant fields $\Psi_i^I$
for a given model and we checked that they and their covariant
adjoints obey local braid relations.  Then we can apply covariant
multiplication and conjugation to obtain arbitrary composites.
No matter how fancy they are, they will always obey local braid
relations. Since  $(\e \o id) R = (id \o \e)R = e$,
it follows in particular that invariant
composites are local observables, i.e. they commute
with all fields in the theory which are localized at spacelike
distance (provided that $\omega^{0I} = \omega^{I0}=1$ which holds
at least for every solution of  (\ref{omegaprop})).

We mentioned that local braid relations are expected to give rise
to representations of the braid group $B_n$ on the space
of n--particle excitations.
When $\R^{IJ>}$ is given
by eq. (\ref{Rmatrix}), this is indeed the case.
Consider states which are created from the ground state
$\vac $ by application of a n-fold product of
field operators $\Psi^{I_k}_{i_k}(x_k,t)$.
Suppose that $x_1<x_2<...<x_n$, for instance.
Operators $\varsigma_k
,( k=1 \dots n-1)$ should act on such states
by interchange of field
operators $\Psi^{I_k}_{i_{k}}$ and $\Psi^{I_{k+1}}_{i_{k+1}}$.
\be
\varsigma^n_k
\Psi^{I_1}_{i_1}\dots \Psi^{I_k}_{i_{k}}
\Psi^{I_{k+1}}_{i_{k+1}} \dots \Psi^{I_n}_{i_n} \vac
 =  \Psi^{I_1}_{i_1}\dots \Psi^{I_{k+1}}_{i_{k+1}}
\Psi^{I_k}_{i_k} \dots \Psi^{I_n}_{i_n}\vac\ \ .
\label{braidinqft}
\ee
Using (\ref{LBR}) and (\ref{Rmatrix}) one verifies by a
 short calculation that the action of
$\varsigma^n_k$ is obtained from the maps
$\sigma^n_k:\G^{*\o n} \mapsto \G^{*\o n}$
introduced in theorem \ref{braidinqqg}.
Since the latter obey Artin relations (\ref{Artinu}) and
the additional factors $\omega^{IJ}$ cancel
in the calculations, eq. (\ref{braidinqft})
defines a representation of the braid group
as it was announced.

\subsection{Quantum symmetry and local braid relations in $\B$ }

When the algebra of field operators $\B$ with
quantum symmetry $\G^*$ was discussed in  section 3.3, we did
not consider locality properties of field operators.
We will demonstrate now that the field operators $\Psi^J_m(\rho_J)$
constructed in section 3.3 satisfy local braid
relations among themselves and with their covariant adjoints.
We need no extra conditions on the
the quantum symmetry $\G^*$ than those
stated there, namely that the fusion rules of the quantum field
theory are in agreement with the selection rules
of the bi-*-algebra $\G^*$ with antipode.
A re-associator $\vp$ and a $R$-element $R$ are constructed from the
fusion- and braiding matrix of the quantum field theory.
In agreement with the general
discussion, they determine the matrix $\R^{IJ}_{kl,ij}$
according to formula
(\ref{Rmatrix}).

Our first aim is to construct appropriate elements
$\vp,R$. We use the same
notations as above. In section 3.3 we
supposed that the dimensions {\it dim\/}$\C(IJ \Y K)$
coincide with the fusion rules $N^{IJ}_K$
defined by the superselection structure.
The latter are commutative and associative in the
sense of eq. (\ref{comass}) so that we are now
in the position to apply theorem
\ref{wqqgconstr}. It asserts that every
solution of the polynomial equations (proposition \ref{poly})
associated with the fusion rules $N^{IJ}_K$
yields elements $\vp,R$ which satisfy
the relations in definition \ref{Defquasiass},
\ref{Defquasitri}. In the present situation the
desired solution is furnished by the
fusion- and braiding-
matrices $F,\Omega$ of the quantum field theory
(proposition \ref{poly}).
Formulae (\ref{pathvp}) for
$\vp$ and (\ref{pathR}) for  $R$
take the following explicit form.
\bea
 C(IP \Y L) C(JK \Y P)_{23} \vp^{IJK} &=& \sum_Q \Fus{P}{Q}{I}{J}{L}{K}^*
 C(QK \Y L) C(IJ \Y Q)_{12} \ \ ,\label{pathvpphys}\\
 C(IJ \Y K) \hat R^{+IJ}  &=& \omega^{IJ}
                         \Oe{K}{I}{J} C(JI \Y K)\ \ ,
\label{pathRphys}
\eea
where $\omega^{IJ}= \overline{\omega^{JI}}$
denotes an arbitrary solution
of (\ref{omegaprop}) by complex phase factors.
The bi-*-algebra $(\G^*,\D,\e,*,\S)$ with antipode $\S$,
re-associator $\vp$ and $R$-element $R$ is
dual to the quantum field theory in the sense
of Fr\"ohlich \cite{Fro3}.

Let us now turn to the discussion of local
braid relations among the field operators
$\Psi^J_m(\rho_J)$.
The statistics operator
$\e(\rho_J,\rho_K)$ was introduced in
(\ref{epsprop}) as an
element of the observables algebra $\A$.
Thus, the action $\pi(\e(\rho_J,\rho_K))$
on the Hilbert space $\H$ is well defined.
Local braid relations will follow from
\be
\Psi^J_j(\rho_J)  \Psi^K_k(\rho_K) \pi(\e(\rho_J,\rho_K))  =
\omega^{JK} \Psi^K_m(\rho_K) \Psi^J_n(\rho_J) {\R}^{JK>}_{nm,jk}\ \ .
\label{Gbraid}
\ee
with $\R$ given by (\ref{Rmatrix}). To prove this identity we insert
definition (\ref{Fieldop}) for the field operators
$\Psi^I_m(\rho_I)$ into the left
hand side and use the
intertwining relation (\ref{iint}) and
the definition (\ref{Brmat})
of the braiding-matrix $B$.
$$
\Psi^J_j(\rho_J) \Psi^K_k(\rho_K) \pi(\e(\rho_J,\rho_K)) =
\sum i^*_{Ll} T_c \vvert{L}{N}{K} T_d \vvert{N}{I}{J} i_{Ii}
\Br{M}{N}{J}{K}{L}{I}^{cd}_{ab}
\CG{J}{M}{L}{j}{m}{l}^a \CG{K}{I}{M}{k}{i}{m}^b\ \ .
$$
With the help of (\ref{Brmat}) the braiding matrix $B$ can be expressed
in terms of the matrices $F,\Omega$ so that it is possible to apply
the definition (\ref{pathvpphys},\ref{pathRphys}) of
$\vp$ and $\R$. The result is
$$
\Psi^J_j(\rho_J)\Psi^K_k(\rho_K) \pi(\e(\rho_J,\rho_K)) = \omega^{JK}
\sum i^*_{Ll} T_c \vvert{L}{N}{K} T_d \vvert{N}{I}{J} i_{Ii'}
\CG{K}{N}{L}{m}{\nu}{l}^c \CG{J}{I}{N}{n}{i}{\nu}^d \tau^I_{i'i}(
\R^{JK>}_{nm,jk}) \ \ .
$$
This equals the right hand side of the equation (\ref{Gbraid}).
Since the statistics operators were
normalized to 1 for $\rho_J > \rho_K$ (cf. (\ref{epsprop})),
local braid relations
follow from eq. (\ref{Gbraid}).

Local braid relations between field operators
$\Psi^I$ and their covariant adjoints are obtained in two steps.
First it is  checked by a calculation similar to
the previous one that
\be
\Psi^J_j(\rho_J) \pi(\e(\rho_K,\rho_J)) (\R^{KJ>})^{-1}_{mn,kj}
\Psi^{K^*}_k (\rho_K) =   \overline{\omega^{KJ}}
\Psi^{K*}_m(\rho_K) \Psi^{J}_n(\rho_J) \ \ .
\ee
Then the Lemma in appendix A establishes the braid relation we are
looking for.
\be
\Psi^K_k (\rho_K) \br\Psi^J_j(\rho_J)  = \omega^{KJ} \br\Psi^J_n
(\rho_J) \Psi^K_m (\rho_K) (\t^K_{mk} \o \bt^J_{nl} \o\U)
(\vp_{213} R_{12} \vp^{-1}) \ \ .
\label{ncbraid}
\ee
Here $\rho_K > \rho_J$ has been assumed.
Braid relations among conjugate fields
follow from the braid relations (\ref{Gbraid})
and the result (\ref{conjbraid}).

\begin{theo} {\em (Local braid relations in $\B$ )}   \label{braidinF}
Suppose that the fusion rules $N^{IJ}_K$ defined by the
superselection structure of $\A$ coincide with the
multiplicities in the Clebsch Gordon decomposition of an
(otherwise arbitrary) semisimple bi-*-algebra with antipode $\G^*$
(in the sense of (\ref{fundAss})). Then the algebra
of field operators $\B = \B(\O)_{\O \in \K}$ with
with quantum symmetry $ \G^* $ constructed
in section 3.3 has the following properties.
\begin{enumerate}
\item
The field operators
(\ref{Fieldop}) which generate the algebras $\B(\O)$
satisfy local braid relations
\be
\Psi^I_i(\rho_I)  \Psi^J_j(\rho_J)  = \omega^{IJ}
\Psi^J_l(\rho_J) \Psi^I_k(\rho_I) {\R}^{IJ>}_{kl,ij}
\label{GbraidII}   \ \ \ \mbox{ for } \ \ \rho_I > \rho_J \ \ .
\ee
Analogous relations (\ref{ncbraid}) hold between the field
operators and their covariant conjugates.
\item
The operators $\R^{IJ>}_{lk,ij}$ are furnished
by elements  $\vp \in
\G^* \o \G^* \o \G^*$ and $R \in \G^* \o \G^*$ which
satisfy all axioms in definition {\rm \ref{Defquasiass},
\ref{Defquasitri}}
\be
{\R}^{IJ>}_{lk,ij}  =
(\tau^I \o \tau^J \o \U)_{lk,ij} (\vp_{213} R_{12} \vp^{-1})
\ \ \ \ .\ee
\item
The quantum symmetry $\G^*$
is a weak quasi quantum group
with re-associator $\vp$ and $R$-element $R$ so that
braid relations are ``stable''
under covariant multiplication (\ref{CovProd})
and conjugation (\ref{covad}) (in the sense
discussed above).
\end{enumerate}
\end{theo}

Even though this theorem was formulated for quantum
field theories on the Minkowski space and for sectors
which satisfy criterion (\ref{DHR}), analogous results
hold for stringlike localized sectors or
quantum fields in chiral conformal quantum
field theory.

Local braid relations do not exploit all the nice properties
of the field operators $\Psi^J_j(\rho_J)$. It was remarked
in \cite{MSIV} that they also satisfy operator product expansions.
\begin{theo} \label{OPE}
             {\em ($\G^*$-covariant operator product expansions)}
The field operators (\ref{Fieldop}) satisfy
\be
(\Psi^J(\rho_J) \ti \Psi^K(\rho_K))_{jk}
 =\sum \Psi^M_m (\rho_M) T_b \vvert{M}{J}{K} \CG{J}{K}{M}{j}{k}{m}^b
 \ \ .
\ee
This holds for any choice of the morphism $\rho_M$ within the
class $[ \rho_M ]$.
\end{theo}

{\sc Proof:}
We use the definition (\ref{Fieldop}) of the field operators and
(\ref{pathvpphys}) of $\vp$ to obtain the first equality in
\ba
(\Psi^J(\rho_J) \ti \Psi^K(\rho_K))_{jk}
& =& \sum i^*_{Ll} T_c \vvert{L}{N}{J} T_d \vvert{N}{I}{K} i_{Ii}
(\Fus {N}{M}{J}{K}{L}{I}^*)^{cd}_{ab}
\CG{M}{I}{L}{m}{i}{l}^a \CG{J}{K}{M}{j}{k}{m}^b\ \nn\\
& = &
 \sum i^*_{Ll} T_a \vvert{L}{I}{M} \rho_I(T_b \vvert{M}{J}{K}) i_{Ii}
\CG{M}{I}{L}{m}{i}{l}^a \CG{J}{K}{M}{j}{k}{m}^b\ \nn \\
& = &\sum \Psi^M_m (\rho_M) T_b \vvert{M}{J}{K} \CG{J}{K}{M}{j}{k}{m}^b\ \ .
\nn
\ea
Application of
relation (\ref{Fusdef}) leads to the  second line where we finally
insert the definition (\ref{Fieldop}) again to obtain the result.

Operator product expansions on the vacuum can be used to convert
vacuum expectation values of products of field operators into
vacuum expectation values of observables. In contrast
with \cite{MSII}, the relations (\ref{OPE}) hold on the whole
Hilbert space $\H$.

In the case of quantum field theories with permutation
group statistics, Doplicher and Roberts have established
the existence of a group algebra $\G^*$ which satisfies
the assumption of theorem \ref{braidinF}. Moreover, they found
that the elements $R$,$\vp$ determined by
(\ref{pathvpphys},\ref{pathRphys}) become trivial, when
the phases $\omega^{IJ}$ are fixed to be $-1$ if
$I,J$ label sectors with para-Fermi statistics and
$+1$ otherwise. Triviality of $R$,$\vp$ means
$R = e\o e $ and $\vp = e \o e \o e $ so that we recover
ordinary Bose-/Fermi (anti-) commutation relations from
equation (\ref{GbraidII}) and the covariant conjugation
coincides with taking adjoints.

\section{The field algebra $\F$} \setcounter{equation}{0}

Field operators $\Psi^J_m$ and  the representation operators
$\U(\xi)$ generate the associative *-algebras
$\B(\O) \subset B(\H)$.
In quantum field theories with permutation group statistics,
one is used to work with field algebras $\F = \F(\O)_{\O \in \K}$
which are generated by fields localized in $\O$.
The (group) symmetry acts on the algebras $\F(\O)$ such
that invariants under this action are local, i.e. commute with
all spacelike separated fields of the theory.
In general, invariant elements within the algebra $\B(\O)$
will not have this property, even if they can be expressed
purely in terms of field operators localized in $\O$ without
any factors $\U(\xi)$.
When $\vp$ is nontrivial there is no way to conclude that such
invariant operators are local with respect to the other fields.
Instead these invariants are products of local fields and
operators $\U(\zeta)$ with $\zeta$ in the center of $\G^*$
\footnote{I thank Karl-Henning Rehren for this
remark}. In other words, the associative algebra $\B(\O)$ is
too large. It is
the factor $\U(\zeta)$ which destroys locality
(unless $\zeta = 1$). Nevertheless
there is a good substitute for the family $\F(\O)$ even
in the case of braid group statistics.Its definition
involves the covariant product $\ti$ which was introduced
in definition \ref{COVPF}.

\begin{defn} {\em (field algebra)}
A net of (not necessarily associative) algebras $\F(\O)_{\O \in \K}$
with conjugation $\overline{\g}$  is called {\em field algebra} $\F$
of the model, if
there is a linear injective map $\pi: \F \mapsto \B(\H)$  and a
set of generators $\psi^J_j$ such that $ \pi(\psi^I_i) = \Psi^I_i$
, $\pi(\br\psi^I_i) = \br\Psi^I_i$  and
\be
\pi(\psi^{I_1}_{i_1} \cdot( \dots \cdot (\psi^{I_{n-1}}_{i_{n-1}} \cdot
\psi^{I_n}_{i_n}) \dots)) =
(\Psi^{I_1} \ti ( \dots \ti (\Psi^{I_{n-1}} \ti
\Psi^{I_n}) \dots))_{i_1 \dots i_{n-1}i_n} \ \ .
\ee
We say that $\pi$ is a {\em representation of the field algebra $\F$}.
\end{defn}

This definition means that the field operators $\Psi^J_m$
faithfully represent the elements $\psi^J_m$ of
the field algebra as linear operators on $\H$.
We will see below that the product $\cdot$
in $\F(\O)$ inherits its properties from the covariant product $\ti$.
In particular it is non-associative in general. The representation
operators $\Psi^J_m$ can be multiplied with the ordinary associative operator
product. The non-associative $\cdot$ product in $\F(\O)$ and the associative
operator product are related indirectly via the covariant product.
Actually, this situation is quite familiar from the representation
theory of Lie algebras.
The non-associative Lie bracket serves as a perfect
analogue of the $\cdot$ product on $\F$. Within a representation
of a Lie algebra, representation operators can be multiplied
using the associative product of linear maps. The
commutator of linear maps expresses the non-associative
Lie product in terms of the operator product.

Given the set of representation operators $\Psi^J_m$
and an appropriate element $\vp$ to define their covariant products,
the field algebra $\F$ is unique and can be constructed
explicitly. In a first step
we define a family of  {\em field sets}
$F(\O) \subset \B(\H)$ to be the linear span of components of
covariant multiplets obtained as a covariant product of the
fundamental field operators. In other words, elements in $F(\O)$
are linear combinations of
\be
\Psi^{I_1, \dots I_{n-1},I_n}_{i_1, \dots i_{n-1},i_n} \equiv
(\Psi^{I_1} \ti ( \dots \ti (\Psi^{I_{n-1}} \ti
\Psi^{I_n}) \dots))_{i_1 \dots i_{n-1}i_n} \ \ ,\label{linspan}
\ee
where all field operators are localized in $\O$.
Obviously, the operator product of two elements in $F(\O)$
will not be in $F(\O)$ in general. To obtain a product $\cdot$ on
$F(\O)$ we make use of the covariant product (\ref{CovProd}). The definition
\be
\Psi^{I_1, \dots I_n}_{i_1, \dots i_n} \ \cdot\
\Psi^{J_1, \dots J_m}_{j_1, \dots j_m}
= \Psi^{I_1, \dots I_n,J_1, \dots J_m}_{i_1,\dots i_n,j_1, \dots j_m} \ \ .
\nn \ee
extends by linearity to $F(\O)$ and furnishes a product on the
field sets $F(\O)$. It follows from relation (\ref{conjprod})
that the covariant conjugation maps the field sets into itself
and hence restricts to a conjugation of the field algebra.
This concludes the construction of $\F$.

Next we introduce an action of the quantum symmetry $\G^*$ on $\F(\O)$, i.e.
we define a linear map $\xi: \F(\O) \mapsto \F(\O)$ for every
element $\xi\in\G^*$ (in an abuse of notation we denote elements in $\G^*$
and the associated linear maps by the same letter $\xi$).
\ba                & &
\xi(\psi^{I_1}_{i_1} \cdot( \dots \cdot (\psi^{I_{n-1}}_{i_{n-1}} \cdot
\psi^{I_n}_{i_n}) \dots)) \nn \\ &\equiv&
(\psi^{I_1}_{k_1} \cdot( \dots \cdot (\psi^{I_{n-1}}_{k_{n-1}} \cdot
\psi^{I_n}_{k_n}) \dots))
(\t^{I_1}_{k_1i_1} \bo ( \dots \bo (\t^{I_{n-1}}_{k_{n-1}i_{n-1}} \bo
\t^{I_n}_{k_n i_n}) \dots)) (\xi)\ \ .
\ea
A tuple $(\psi_\a)$ of elements $\psi_\a \in \F(\O)$ is said to transform
covariantly according to the representation $\t$ of $\G^*$ if
\be
\xi (\psi_\a) = \psi_\b \t_{\b \a} (\xi) \ \ ,
\ee
for all $\xi \in \G^*$. $\phi \in \F$ is invariant if
it transforms according to the trivial representation
$\e$ of $\G^*$, i.e. if $\xi(\phi) = \phi \e(\xi)$.

 The following theorem collects all the important properties of
the field algebra $\F = \F(\O)_{\O \in \K}$.

\begin{theo}
{\em (properties of the field algebra $\F$ )}
\begin{enumerate}
\item
    The algebras $\F(\O)$ are quasi--associative in the sense that the
    product $\psi^{J_n}_{j_n} \cdots \psi^{J_1}_{j_1}$
    with arbitrary specification
    of the position of brackets can be written as a complex linear
    combination of products
    $\psi^{J_n}_{k_n} \cdots \psi^{J_1}_{k_1}$ with any other
    specification of brackets. Re-association
    is performed with the help of the formulae
   \ba
    ((\psi_{\a} \cdot \psi'_{\b}) \cdot \psi''_{\c}) & = &
    (\psi_{\d} \cdot (\psi'_{\e} \cdot \psi''_{\kappa}))( \tau_{\a\d} \o
    \tau'_{\b\e} \o \tau''_{\c\kappa}) (\vp )\ ,\\
    (\psi_{\d} \cdot (\psi'_{\e} \cdot \psi''_{\kappa})) & = &
    ((\psi_{\a} \cdot \psi'_{\b} ) \cdot \psi''_{\c} )
    (\tau_{\d\a} \o \tau'_{\e\b} \o \tau''_{\kappa\c}) (\vp ^{-1}) \ .
   \ea
    They are valid if $(\psi_\a),(\psi'_\b),(\psi''_\c)$ transform
    according to representations $\tau,\tau',\tau''$ of $\G^*$.
\item
    Quantum fields obey local braid relations, i.e. if $\psi_\a \in \F(O_1)$,
    $\psi'_\b \in \F(O_2)$ transform according to irreducible
    representations $\t \cong \t^J$ and  $\t' \cong \t^K$
    of $\G^*$,
   \be
    \psi_\a \cdot \psi'_\b = \omega^{JK} \psi'_\c \cdot \psi_\d
    (\tau_{\d\a} \o \tau'_{\c\b})(R)
   \ee
    whenever $\O_1 > \O_2$.
\item
    $\xi \in \G^*$ acts on the algebras $\F(\O)$ as a generalized
    derivation, i.e.
   \ba
    \xi (\psi \cdot \psi' ) & = & \sum_\s \xi^1_\s
    (\psi)\cdot \xi^2_\s(\psi')
   \ea
    for arbitrary $\psi,\psi' \in \F$.
\item
    The conjugation $\overline{\g}$ on $\F(\O)$ is involutive,
    $\br{\br\psi} = \psi$, and satisfies
   \be
     \overline{ (\psi \cdot \psi')}_{\a\b} =
     (\br\psi' \cdot \br\psi)_{\d\c}
     (\bt'_{\d\b} \o \bt_{\c\a})(f^*) \ \ .
   \ee
    where $f$ denotes the element (\ref{fel}) and $\psi,\psi'$
    are again assumed to transform covariantly according to
    the representations $\t, \t'$.
\item
    The algebras $\A(\O)$ consist of all  invariants in $\F(\O)$, i.e.
   \be
    \A(\O) \cong \{ \psi \in \F(\O)  \Y  \xi(\psi) = \psi \e(\xi) \} \ \ .
   \ee
    On invariants $\phi \in \A(\O)$ the covariant conjugation restricts
    to the adjoint $*$, $\br\phi = \phi^*$.
\end{enumerate}
\end{theo}

Observe that the version of local braid relations we used
in this theorem implies local braid relations for arbitrary pairs
of generating fields $\psi^I_i$ and their conjugates $\br\psi^J_j$.
If the quantum fields $\psi_\a$ transforms according to a reducible
representation $\t$, Clebsch Gordon intertwiners corresponding
to the irreducible subrepresentations  of $\t$
furnish a linear decomposition of the multiplet $(\psi_\a)$.
The relations $2.$ apply -- by construction -- to the
individual summands in this linear combination so that
local braid relations for arbitrary quantum
fields $\psi_\a$ can be worked out.

Let us just remark at the end that operator product expansions
of the form (\ref{OPE}) give rise to operator product expansion
in the field algebra.
\be
\psi^J_m(\rho_J) \cdot \psi^K_n(\rho_K)
 = \sum \psi^M_m (\rho_M) T_b \vvert{M}{J}{K} \CG{J}{K}{M}{j}{k}{m}^b\ \ .
\ee

\section{Discussion and Outlook}  \setcounter{equation}{0}

For the construction of the field algebra $\F$ with
weak quasi quantum group symmetry, every bi-*-algebra
with antipode $\G^*$
can be admitted, provided the multiplicities in
the Clebsch Gordon decomposition coincide with
the fusion rules of quantum field theory.
This selection criterion for $\G^*$ as well as the
explicit expressions for the re-associator and
the $R$-matrix were completely determined by
the observables $\A$ of the model.

In \cite{MSI} we used algebraic methods to
construct a field algebra with quantum symmetry
for the critical chiral Ising model. Instead of
$C^*$-algebras we used the
Virasoro algebra Vir$_{c=1/2}$
as a chiral Lie algebra  Lie$\A$
of observables. This Lie algebra admits
three inequivalent unitary irreducible
positive energy representations $\pi_J$
in the Hilbert spaces $\H^J,J=0,\frac12,1$.
To construct suitable endomorphisms,
we had to enlarge Vir$_{c=1/2}$ to
a Lie algebra Lie$\bar \A$. All unitary
irreducible positive energy representations
$\pi_J$ of the Virasoro algebra Vir$_{c=1/2}$
on $\H^J$ extend to representations of Lie$\bar \A$
in the same Hilbert space $\H^J$.
A complete set of endomorphisms $\rho_J$
of Lie$\bar \A$ to reach all the sectors
was found explicitly. The fusion rules
were calculated from the explicit expressions.
\bea
[\rho_{1/2}\circ \rho_{1/2} ] & = & [\rho_0]+[\rho_1]\ \ ,\nn\\ \,
[\rho_{1/2}\circ \rho_1] &=& [\rho_1\circ \rho_{1/2}] =
[\rho_{1/2}]\ \ ,\label{Isifus} \\   \,
[\rho_1 \circ \rho_1] & = & [\rho_0]
\nn \eea
and $[\rho_0 \circ \rho_J] = [\rho_J \circ \rho_0]= [\rho_J]$
for all $J=1,\frac12,1$. The matrices $N^{IJ}_K$ can
be read off.

It was suggested in \cite{MSIII,MSIV} to regard a ``truncated''
version of the quantum group algebra
$U_q(sl_2)$ as quantum symmetry of the chiral Ising model.
As we have seen above, the critical Ising model
(as any other rational model) admits
infinitely many quantum symmetries. It is important
to remark that they are all truncated
(in particular they cannot be group symmetries).
Without truncation
the dimensions $\delta_J$ of irreducible
representations $\t^J$ have to satisfy
\be \delta_I \delta_J = \sum_K N^{IJ}_K \delta_K \label{Fuseq}\ \ .\ee
When $N^{IJ}_K$ are the fusion rules of a quantum field theory
with permutation group statistics, an
integer solution of this equation is furnished
by the ``statistics dimensions'' $d_J$ of the
superselection sectors \cite{DHR2}.
The  set of statistics dimensions $d_J$ gives at the same time
the dimensions of representations of the
symmetry groups constructed by Doplicher
and Roberts. On the simple level of eq. (\ref{Fuseq})
differences with the situation in low dimensional quantum
field theory show up already.
As one can easily check for the fusion
rules of the chiral critical Ising model,
positive integer solutions of (\ref{Fuseq})
do not exist. Whenever this happens
it excludes the
possibility of non-truncated quantum symmetries.
If one admits for truncation, the condition (\ref{Fuseq})
on the dimensions
$\delta_J$ of irreducible representations
becomes an inequality, which has
an infinite number of solutions so that there
is a priori an enormous freedom in the construction
of quantum symmetries (and consequently in the construction
of field algebras).

If we fix a particular integer solution of the
inequality $\d_I \d_J \geq \sum N^{IJ}_K \d_K$, the
remaining freedom is reduced to a ``twist'' in the
sense of \cite{Dri2}. As H. Rehren remarked in \cite{Rehtalk},
this corresponds to the possibility of Klein
transformations of field operators.

It should be mentioned at the end that
soliton sectors in massive two--dimensional
quantum field theory do
not fit into the present
theory of superselection structure.
The usual analysis applies only to a special
class of models which was discussed by
Fr\"ohlich \cite{Fro1}.
Inspired by the properties of classical
multi-soliton solutions, Fredenhagen proposed
that in generic situations, soliton sectors can only
be composed if they ``fit together''
\cite{Fre3,FreKar}.
The structure of possible ``quantum symmetries''
is not known for these more general cases,
but will probably be quite different
from the quantum symmetries treated here.

Validity of the analysis in this paper is also restricted
to finite statistics and does not extend to quantum field
theories with infinite statistics. Models
of such theories exist \cite{Fre4}.

\section{Appendix A: Properties of covariant conjugation}

This appendix is devoted to the properties of the covariant
conjugation which was introduced in section 2. A list of
properties is given in section 5.1.. Their proofs will be
sketched below.

As we remarked before, the covariance law can be used to shift
operators $\U(\xi)$ from the left to right of the fields $\Psi^I_i$.
Yet we did not discuss how to achieve the reverse, i.e.
move elements from right to left. It is precisely the existence
of an intertwiner $\vp$ satisfying (\ref{vpint}) which allows to do
this. Let us introduce the element $w = \sum \vp_\s^2 \S^{-1} (\vp^1_\s\b)
\o \vp^3_\s$ as before and define a new multiplet $\pr{i}\Psi^I$
according to
\be
\pr{i}\Psi^I (x,t) \equiv \Psi^I_j (x,t)
(\t^I_{ji} \o \U) (w) \ \ .  \label{rlcov}
\ee
This new tuple has a ``good'' transformation behaviour, namely
\be \pr{i}\Psi^I \U(\xi) = (\tt^I_{ij} \bo \U) (\xi) \pr{j}\Psi^I \ \ ,
\label{lcov} \ee
where $\tt(\xi) \equiv \ ^t\t^I(\S^{-1} (\xi))$ for all $\xi \in \G^*$
and all representations $\t$ of $\G^*$. We will often refer to
transformation laws of the form (\ref{lcov}) as ``left covariance''
to distinguish them from the (right-) covariance (\ref{covtrans}).
Relation (\ref{lcov}) is actually a simple consequence of the
intertwining property (\ref{vpint}) and the relations (\ref{antip}).

If the element $\vp$ is the re-associator of a weak quasi quantum
group (i.e. $\vp$ satisfies all the relations in section 4.1) the
map (\ref{rlcov}) from (right-) covariant to left-covariant
multiplets has a number of useful properties.  They are summarized
in the following proposition.

\begin{prop} {\em (properties of def. (\ref{rlcov}))}
\label{proprl}
Assuming that the  quantum symmetry $\G^*$
is a weak quasi quantum group one can show
\begin{enumerate}
\item
  the map (\ref{rlcov}) from right- to left-covariant elements
  has an inverse,
 \be
   \Psi^I_i (x,t) =
   (\tt^I_{ji} \o \U) (v) \pr{i}\Psi^I (x,t)\ \ . \label{lrcov}
 \ee
 Here $v \in \G^* \o \G^*$ is defined by
 $v = \sum \S(\phi^2_\s) \a \phi^1_\s \o \phi^3_\s$ with
 $\phi^i_\s$ given through the expansion of $\phi = \vp^{-1}$.
\item
 the passage from right- to left covariant elements is consistent
 with local braid relations in the following sense. Suppose that
 $\Psi^I_i, \Psi^J_j$ satisfy local braid relations (\ref{Rtilde}).
 Then
  \be
   \pr{i} \Psi^I \pr{j} \Psi^J = \omega^{IJ}
   (\tt^I_{ki}  \o \tt^J_{lj} \o \U)(\vp_{213} R_{12} \vp^{-1})
   \pr{l} \Psi^J \pr{k} \Psi^I   \ \ .
   \label{lcovbraid}
  \ee
\item
 on covariant products, (\ref{rlcov}) acts according to
  \be
   (\Psi^J \ti \Psi^I)_{mn} ((\t^J \bo \t^I)_{mnji} \o \U )(w) =
   (\tt^I_{ik} \o \tt^J_{jl} \o \U) ( f_{12} \vp^{-1})
   \pr{l} \Psi^J \pr{k} \Psi^I
  \ee
 where $f \in \G^* \o \G^*$ is the element (\ref{fel}).
 \end{enumerate}
\end{prop}

{\sc Proof:} A detailed proof of this proposition is beyond the
scope of this text. We start with the first item and make only some
remarks on the two others.
The relation (\ref{lrcov}) is a straight forward
application of the pentagon equation (\ref{phiphi}).
The latter can be restated as
$$
(id \o id \o \D) (\vp^{-1}) (e \o \vp) =  (\D \o id \o id)(\vp)
( \vp^{-1} \o e) (id \o \D \o id) (\vp^{-1}) \ \ .
$$
Using the properties (\ref{antip}) of the antipode one derives
$$
(id \o \D)(v) \vp = \sum (\S(\phi^1_\t \phi^1_\s) \a \phi^2_\t
                  \o \phi^3_\t \o \phi^3_\s) (\D(\phi^2_\s) \o e)
$$
and then after a similar step (with $v = \sum v^1_\s \o v^2_\s$)
$$
\sum \D(v^2_\s)  w  (\S^{-1}(v^1_\s) \o e)
   = \sum (\phi^3_\s \S^{-1}(\a\phi^2_\s\b)
  \phi^1_\s \o e)\D(e) = \D(e)\ \ ,
$$
where the last equality follows from relation (\ref{antonvp}).
This equation can be exploited in the following calculation.
\bea
  (\tt^I_{ji} \o \U) (v) \pr{i}\Psi^I & = & \Psi^I_k (\tt^I_{ji} \o
 \t^I_{kl} \o \U)((id \o \D)(v)) (\t^I_{li} \o \U)(w) \nn\\
 & = & \Psi^I_k (\t^I_{kj} \o \U)(\D(e)) = \Psi^I_j \ \ .\nn
\eea

We prove the second item by iteration of arguments leading to
the following lemma. \\[4mm]
{\bf Lemma:} \it Suppose that $\Psi^I_i, \Psi^J_j$ satisfy local
 braid relations (\ref{Rtilde}). Then
  \be
   \omega^{IJ} \pr{j} \Psi^J \ \Psi^I_i =
   \Psi^I_k \  (\t^I_{ki}  \o \tt^J_{jl} \o \U)(\R)
   \pr{l} \Psi^J \ \ ,
  \ee
 with $\R = \vp_{213} R_{12} \vp^{-1}$.  \rm \\[4mm]
{\sc Proof: }  From the defining relations of a weak quasi quantum group
one can derive the following {\em generalized quasi-triangularity}
\be
 \R^{-1}_{134}
[ (id \o id \o \D) (\vp)]_{2314} (id \o \D \o id)(\R)
=  (id \o id \o \D)(\R) (e \o \vp)  \ \ .        \nn
\ee
The meaning of the lower indices has been explained in definition
\ref{Defquasitri}. This equation is a good starting point to
obtain
\be
(\varrho^1_\s \o \D(\varrho^3_\s))
(e \o w) (e \o \S^{-1}(\varrho^2_\s) \o e) =
\R^{-1} [ (id \o \D)(w)]_{213} \ \ ,
\nn \ee
where $\varrho^i_\s$ are the components of $\R$. As a consequence
\bea
   \Psi^I_k \  (\t^I_{ki}  \o \tt^J_{jl} \o \U)(\R)
   \pr{l} \Psi^J
   & = & \Psi^I_k \ \Psi^J_l \ (\t^I_{ki} \o \t^J_{lj}
    \o \U)( R^{-1} [(id \o \D)(w)]_{213} ) \nn \\
   & = & \omega^{IJ} \ \Psi^J_n\  \Psi^I_m (\t^I_{mi} \o \t^J_{nj} \o \U)
    ([ (id \o \D)(w)]_{213}) \nn \\
   & = & \omega^{IJ}\  \Psi^J_n (\t^J_{nj} \o \U)(w) \ \Psi^I_i \nn\\
   & = & \omega^{IJ} \pr{j} \Psi^J \ \Psi^I_i \ \ . \nn
\eea
This is the proposed equation.

{\sc Proof of proposition \ref{proprl}}{\em (continued)}:
Things become really cumbersome when we come to the third item.
Here is a rough sketch of the proof. One has to insert the definition
of left covariant multiplets into the left hand side of the
equation. Next the covariance law is applied to move all factors
which involve elements in the symmetry algebra $\G^*$ to the right.
Once this is done, the resulting expression is rewritten with the help
of the pentagon equation (\ref{phiphi}) and the properties (\ref{antip})
of the antipode. One has to use (\ref{phiphi})
three times until everything simplifies as a consequence of
the formulae (\ref{fint} f.) for the element $f$.
This completes the proof of the proposition.

The covariant conjugation (\ref{covad}) introduced in
section 2 amounts to the prescription
\be
\br\Psi^I_i \equiv (\pr{i} \Psi^{I})^*\ \ .
\ee
The transformation law (\ref{adtrans}) of
$\br\Psi^I$ is a direct consequence
of relation (\ref{lcov}) above. All properties of the covariant
conjugation listed in section 5.1 follow from proposition
\ref{proprl} (notice that $w^* = v$). \\[2cm]
%
%
%
{\bf Acknowledgements:} It is a pleasure to thank Gerhard
Mack for many helpful and encouraging conversations and for providing
the final stimulus to finish this work. I also profited a lot from
discussions with Henning Rehren, Klaus Fredenhagen
and Kornel Szlachanyi. The warm hospitality at
the Erwin Schr\"odinger Institute (where many of the discussions
took place) and in particular of Harald
Grosse is gratefully acknowledged. Im am indebted to Arthur
Jaffe for the opportunity to continue my work in his group.
This project was partially supported by  the NSF grant No.
PHY-91-20626.

\end{document}